\documentclass[10pt,a4paper]{extarticle}
\usepackage{asymmetry}
\usetikzlibrary{fit}

\usepackage{etoolbox}
\pretocmd{\section}{\Needspace{7\baselineskip}}{}{\PackageWarning{asymmetry}{section guard not applied}}

\newcolumntype{Y}{>{\raggedright\arraybackslash}X}
\newenvironment{claimnotes}
  {\par\vspace{2.5pt}\begin{minipage}{\linewidth}\footnotesize
   {\sffamily\bfseries\color{ACslate}Claim notes.}\enspace
   \setlength{\parskip}{1pt}}
  {\end{minipage}}

\begin{document}
\thispagestyle{firstpage}
\vspace{2pt}

\ACmaketitle%
  {Asymmetry PRISM: A CPU/GPU Portfolio Optimization Engine}%
  {for Deadline-Bounded Institutional Rebalancing}

\ACauthorline

\begin{abstractbox}
\begin{center}{\sffamily\bfseries\color{ACnavy}Abstract}\end{center}%
\vspace{-2pt}
\noindent{\color{ACline}\rule{\linewidth}{0.4pt}}\par
\vspace{4pt}
\small
Institutional rebalancing is a batched optimization workload with a hard
operating deadline: hundreds of accounts need new weights under budget,
turnover, exposure, exclusion, and tax-aware controls before trading can
proceed. This paper evaluates Asymmetry PRISM, a CPU/GPU portfolio
optimization engine, through a public evaluation boundary; problem data in,
and returned weights, status codes, timings, memory class, external
feasibility diagnostics, eligible objective comparisons, and audit records
out. Within that boundary, the evaluation protocol fixes hardware and software
versions, declares timing
lanes, separates cold single calls from repeated workloads, and admits
objective-gap claims only where an eligible reference solver completed. On
completed multi-solver rows from $N{=}100$ to $N{=}2{,}000$, Asymmetry PRISM-CPU is
$4.5\times$ to $24.1\times$ faster than the fastest completed reference row in
the same lane. In the production queue study, Asymmetry PRISM-GPU completes 500/500
accounts over a 10,000-instrument universe in 109.5\,s within a declared
25-minute operating window, with zero missed deadlines and an audit record for
every solve; the recorded OSQP queue baseline completes 4/500. On an
operationally constrained real-data suite (tax-motivated transition
penalties, restriction caps, turnover controls, batches), Asymmetry PRISM clears constrained
solves $3.4\times$ to $126.7\times$ faster than the best completing
incumbent at certified-equal objectives, and the GPU route widens to $8.8\times$ over
the CPU route at $N{=}384{,}800$. Rows without a completed reference are
reported as feasibility, timing, memory, and failure-status evidence.
\end{abstractbox}

\vspace{5pt}
{\small\noindent{\sffamily\bfseries\color{ACnavy}Evaluation artifacts}\enspace
\href{https://github.com/AsymmetryComputing/prism-public-evaluation}%
{\faGithub\,\texttt{github.com/AsymmetryComputing/prism-public-evaluation}}
\ (release \texttt{v1.0.0}): result tables, evidence ledger, and external
feasibility and residual checks.}

\vspace{2pt}
{\small\noindent{\sffamily\bfseries\color{ACnavy}Keywords}\enspace
portfolio optimization, GPU-native execution,
deadline-bounded optimization, institutional rebalancing, solver benchmarking,
audit-reproducible computation.}

\vspace{5pt}

\section{Introduction}
\label{sec:intro}

In production rebalancing, the unit of work is a book of accounts, not an
isolated optimization solve. Each account carries budget, position, exposure,
turnover, exclusion, cash-management, and tax-aware controls, and the run
succeeds only if the required account set is solved, validated, and recorded
before the trading workflow advances. Mathematical quality matters. Elapsed
time matters equally: a portfolio that arrives after the trading window is
operationally unusable, even if a solver can later certify a better answer.

The operating questions are therefore systems questions. Does every account
return a tradable weight vector before the deadline? Do independent checks
confirm feasibility? Are failures classified rather than hidden? Does an audit
record exist for each decision? A scalar objective value answers none of these
on its own, so deadline completion, external feasibility, and auditability are
treated as first-class metrics, alongside objective quality where it can be
certified.

\begin{figure}[H]
\centering
\resizebox{0.985\textwidth}{!}{%
\begin{tikzpicture}[
  yscale=1.34,
  font=\footnotesize,
  box/.style={rectangle, rounded corners=3pt, draw=ACslate!70, fill=white,
              align=center, inner sep=6pt, minimum height=0.95cm},
  gen/.style={box, fill=ACfill, draw=ACnavy!60, thick},
  enc/.style={box, fill=ACblue!6, draw=ACblue!55},
  comp/.style={box, fill=white, draw=ACslate!60, minimum width=3.55cm},
  prism/.style={box, fill=ACblue!8, draw=ACnavy!70, thick, minimum width=6.1cm},
  lane/.style={box, fill=ACviolet!6, draw=ACviolet!55, text width=3.05cm,
               minimum width=3.55cm, minimum height=1.35cm},
  evid/.style={box, fill=ACteal!6, draw=ACteal!60},
  gov/.style={box, fill=ACcard, draw=ACslate!70, minimum width=4.9cm,
              minimum height=1.2cm},
  gate/.style={lane, fill=ACamber!8, draw=ACamber!70},
  arr/.style={-{Stealth[length=2.4mm]}, ACslate!85, line width=0.75pt},
  busline/.style={ACslate!85, line width=1.15pt},
  grouplab/.style={font=\scriptsize\itshape, text=ACslate}
]

\node[gen, minimum width=11.6cm] (genbox) at (8.6,15.2)
  {\textbf{Public benchmark workload: batched account rebalancing}\\[1pt]
   {\scriptsize convex problem class Eq.~\eqref{eq:public_problem} per account: budget, box, exposure bands, turnover, policy halfspaces, tax-aware penalty;}\\[0.5pt]
   {\scriptsize real S\&P~500 universe (481 names, 2019--2025) and a disclosed real-calibrated extension; fixed seeds; archived price cache; one shared market deadline}};

\node[enc, minimum width=6.4cm] (tocomp) at (4.0,13.1)
  {\textbf{Identical public instance}, Eq.~\eqref{eq:public_problem}\\[1pt]
   {\scriptsize served to every comparator (same $\bmu, Q$, constraints)}};
\node[enc, minimum width=6.4cm] (toprism) at (13.2,13.1)
  {\textbf{Same instance} $(\bmu, Q, \text{constraints})$\\[1pt]
   {\scriptsize served to Asymmetry PRISM; no reformulation advantage}};

\draw[arr] (genbox.south -| tocomp.north) -- (tocomp.north);
\draw[arr] (genbox.south -| toprism.north) -- (toprism.north);

\node[comp] (osqp)     at (2.15,11.2) {OSQP 1.1.0\\[-1pt]{\scriptsize full-call wall-clock}};
\node[comp] (clarabel) at (5.95,11.2) {Clarabel 0.11.1\\[-1pt]{\scriptsize via CVXPY 1.8.1}};
\node[comp] (scs)      at (2.15,10.0) {SCS\\[-1pt]{\scriptsize via CVXPY 1.8.1}};
\node[comp] (mosek)    at (5.95,10.0) {MOSEK 11.1.11\\[-1pt]{\scriptsize interior point}};
\node[comp, minimum width=7.35cm] (ref) at (4.05,8.8)
  {Anonymized commercial reference\\[-1pt]{\scriptsize objective-gap eligibility only; status, not a published timing row}};

\begin{scope}[on background layer]
\node[draw=ACslate!55, dashed, rounded corners=4pt, fill=ACfill!40,
      fit=(osqp)(clarabel)(scs)(mosek)(ref), inner sep=10pt] (compgroup) {};
\end{scope}

\node[prism] (routes) at (13.2,10.75)
  {\textbf{Measured execution configurations}\\[1pt]
   {\scriptsize Asymmetry PRISM-CPU and Asymmetry PRISM-GPU}\\[0.5pt]
   {\scriptsize evidence build \texttt{512758f}}};
\node[prism] (outputs) at (13.2,9.0)
  {\textbf{Returned observables only}\\[1pt]
   {\scriptsize weights $\widehat{\bw}$, status codes, full-call and device timings,}\\[0.5pt]
   {\scriptsize memory class, feasibility diagnostics, audit record}};
\node[grouplab, text=ACnavy!85] (ifacenote) at (13.2,7.85)
  {only this interface crosses the boundary};

\begin{scope}[on background layer]
\node[draw=ACnavy!75, dashed, thick, rounded corners=4pt, fill=ACblue!4,
      fit=(routes)(outputs)(ifacenote), inner sep=10pt] (prismgroup) {};
\end{scope}

\draw[arr] (tocomp.south) -- (tocomp.south |- compgroup.north);
\draw[arr] (toprism.south) -- (toprism.south |- prismgroup.north);
\draw[arr] (routes.south) -- (outputs.north);

\node[grouplab, fill=white, inner sep=1.5pt, anchor=south west] at ([yshift=3pt]compgroup.north west)
  {claim-bearing comparators (CVXPY model build inside the timed call)};
\node[grouplab, text=ACnavy!80, fill=white, inner sep=1.5pt, anchor=south west] at ([yshift=3pt]prismgroup.north west)
  {Asymmetry PRISM: implementation outside the public artifact};

\coordinate (busL) at (0.6,6.0);
\coordinate (busR) at (16.6,6.0);
\draw[busline] (busL) -- (busR);

\draw[arr] (compgroup.south -| ref.south) -- (ref.south |- busL);
\draw[arr] (prismgroup.south -| outputs.south) -- (outputs.south |- busL);

\node[lane] (l1) at (2.35,4.75)
  {\textbf{Full-call wall-clock}\\[0.5pt]{\scriptsize multi-solver and burden (E1)}};
\node[lane] (l2) at (6.35,4.75)
  {\textbf{Core/device solve}\\[0.5pt]{\scriptsize GPU reported solve (E6)}};
\node[lane] (l3) at (10.35,4.75)
  {\textbf{Queue wall-clock}\\[0.5pt]{\scriptsize throughput, miss rate (E5)}};
\node[gate, text width=3.45cm, minimum width=3.85cm] (d1) at (14.45,4.75)
  {\textbf{Deadline gate}\\[0.5pt]{\scriptsize 25-min operating window,}\\[-1pt]
   {\scriptsize real-data scenarios (E7)}};

\foreach \n in {l1,l2,l3,d1}{\draw[arr] (\n.north |- busL) -- (\n.north);}

\node[align=center, font=\scriptsize\itshape, text=ACslate, text width=8.3cm,
      fill=white, inner sep=2pt] at (8.6,6.78)
  {measurement boundary: full-call wall-clock at the public\\
   solver-call boundary; external KKT-style residuals and\\
   certified objective gap where eligible; fixed status vocabulary};

\node[evid, minimum width=15.8cm] (evidence) at (8.6,3.0)
  {\textbf{Versioned evidence artifacts} (JSON/CSV; runtime-environment block, seeds, per-instance rows, failure counters)\\[1.5pt]
   {\scriptsize \texttt{E1} multi-solver public rows $\cdot$ \texttt{E3} external KKT diagnostics $\cdot$ \texttt{E5} production queue $500{\times}10{,}000$}\\[0.5pt]
   {\scriptsize \texttt{E6} GPU timing-lane separation $\cdot$ \texttt{E7} hard scenarios on real data}};

\foreach \n in {l1,l2,l3,d1}{\draw[arr] (\n.south) -- (\n.south |- evidence.north);}

\node[gov] (consol) at (2.95,1.0)
  {\textbf{Generated, never typed}\\[-1pt]
   {\scriptsize consolidation scripts $\to$ table bodies,}\\[-2.5pt]
   {\scriptsize figures, envelope ranges}};
\node[gov] (contract) at (8.6,1.0)
  {\textbf{Claim contract gate}\\[-1pt]
   {\scriptsize same-lane ratios only; anonymized row}\\[-2.5pt]
   {\scriptsize $=$ objective-gap eligibility, not a timing claim}};
\node[gov] (paper) at (14.25,1.0)
  {\textbf{Paper claims and ledgers}\\[-1pt]
   {\scriptsize result card, tables, eligibility ledger,}\\[-2.5pt]
   {\scriptsize claim-to-evidence matrix}};

\draw[arr] (evidence.south -| consol.north) -- (consol.north);
\draw[arr] (consol.east) -- (contract.west);
\draw[arr] (contract.east) -- (paper.west);

\end{tikzpicture}}%
\vspace{6pt}
\caption{End-to-end evaluation architecture. A public benchmark workload of
batched account rebalances instantiates the convex problem class
Eq.~\eqref{eq:public_problem} per account; the identical instance is served to
every claim-bearing comparator (OSQP, Clarabel, SCS, MOSEK via CVXPY, plus an
anonymized commercial reference used for objective-gap eligibility only) and to
Asymmetry PRISM (dashed boundary). Comparator rows include CVXPY model
construction inside the timed call, strict in Asymmetry PRISM's disfavor. Only
returned observables cross that boundary: weights, status
codes, timings, memory class, feasibility diagnostics, and an audit record.
Every execution passes one measurement boundary (full-call wall-clock,
external KKT-style residuals where eligible, a fixed status vocabulary) and is
routed through exactly one declared lane: the full-call lane (E1), the
core/device-solve lane for the GPU reported-solve interval (E6, never compared
across lanes), the queue wall-clock lane (E5), or the 25-minute deadline gate
of the real-data scenario suite (E7). Lanes append to versioned evidence
artifacts, and the paper consumes those artifacts only through the generation
scripts and the claim contract, which permits same-lane ratios only and
restricts the anonymized reference to objective-gap eligibility.}
\label{fig:architecture}
\end{figure}
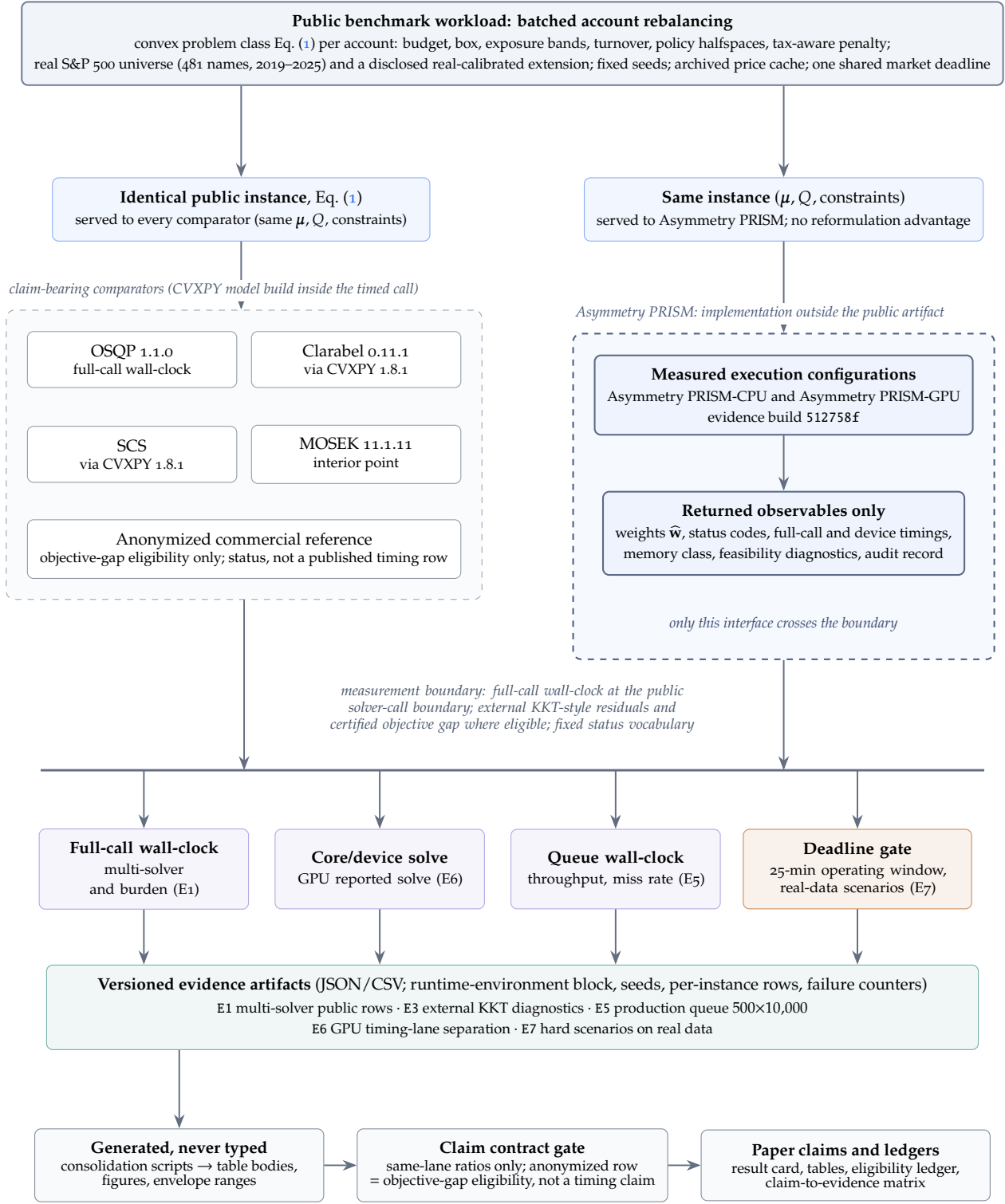

Asymmetry PRISM is a CPU/GPU optimization engine built for this workload.
The paper evaluates it through externally observed input--output behavior
under a fixed, disclosed protocol (Section~\ref{sec:claim_contract}), and
claims only what that evidence supports.
Figure~\ref{fig:architecture} shows the end-to-end evaluation architecture,
from public instances through the measurement boundary and declared timing
lanes to the versioned evidence artifacts and the claim contract that governs
every published number.

\Needspace{9\baselineskip}
\begin{keyresult}
\begin{itemize}[leftmargin=1.5em, itemsep=1.5pt]
  \item \textbf{500/500} accounts completed in the production queue study
        (10,000-instrument universe), with an audit record for every solve.
  \item \textbf{109.5\,s} total queue wall-clock, inside a declared
        25-minute operating window; \textbf{0} missed deadlines.
  \item \textbf{4.5$\times$ to 24.1$\times$} faster than the fastest completed
        reference row on the $N{=}100$ to $N{=}2{,}000$ multi-solver
        benchmarks.
  \item The recorded OSQP queue baseline completes \textbf{4/500} accounts
        under the same protocol.
  \item Real-data scenario suite: \textbf{3.4$\times$ to 126.7$\times$}
        faster than the best completing incumbent at certified-equal
        objectives; the GPU route widens to \textbf{8.8$\times$} over CPU
        at $N{=}384{,}800$.
\end{itemize}
Large-scale rows without a completed reference solver are reported as
feasibility, timing, memory, and failure-status results.
\end{keyresult}

The paper makes four contributions.

\begin{enumerate}[label=\textbf{C\arabic*.}, leftmargin=2.4em]
  \item \textbf{A public evaluation boundary.}
        The benchmark discloses hardware, software versions, timing lanes,
        tolerance gates, failure statuses, and evidence provenance.

  \item \textbf{A public benchmark protocol for recorded eligible baselines.}
        Asymmetry PRISM is compared against recorded eligible baselines under declared
        timing lanes, status rules, and external feasibility checks
        (Section~\ref{sec:protocol}).

  \item \textbf{Separated quality and deadline claims.}
        Objective gaps appear only where a reference solver completed on the
        same public objective. Rows above that boundary report external
        feasibility and deadline outcomes.

  \item \textbf{Production workflow and real-data scenario evidence.}
        The queue benchmark measures account completion, missed-deadline rate,
        p50/p99 timing, and audit-record coverage, and an operationally
        constrained real-data suite stresses tax-motivated transition
        penalties, restriction caps, turnover controls, deadline budgets,
        batch throughput, and scale; this is the form in which institutional
        rebalancing is actually operated.
\end{enumerate}

\section{Related Work and Evaluation Gap}
\label{sec:literature}

This section separates the literature into five roles: portfolio models
define the objective language; personalized and tax-aware work defines the
account-level constraint burden; solver systems define the software comparison
context; benchmarking literature defines the evaluation discipline; and
deadline-bounded computation, together with pre-trade control requirements,
explains why completion before an operating window is a first-class metric.

\subsection{Portfolio Construction Theory}

The Markowitz framework formalized the tradeoff between expected return and
risk \citep{markowitz1952}. Sharpe's capital-asset-pricing model and the
Fama--French factor evidence established systematic risk models as the central
portfolio language \citep{sharpe1964capital,fama1993common}. Michaud documented
the instability that arises when estimated inputs feed optimizers directly
\citep{michaud1989markowitz}; Black--Litterman combined equilibrium structure
with investor views \citep{black1992global}; and Ledoit--Wolf shrinkage remains
the core reference for high-dimensional covariance estimation
\citep{ledoitwolf2004,ledoit2017nonlinear}.
These models define the objective language. They do not, by themselves, define
the operational workload faced when many constrained accounts must be updated
under a deadline.

\subsection{Personalized and Tax-Aware Implementation}

Tax-managed investing, personalized indexing, and direct indexing introduce
account-specific restrictions, turnover costs, tax lots, exclusion lists, and
client-level policy controls
\citep{stein2017,khang2022,moehle2021taxaware,moehle2023lcso}. Cost-aware
portfolio work places transaction costs inside the optimization problem rather
than as an after-the-fact adjustment \citep{fan2025costaware}, and recent
evidence shows that the timing and market footprint of rebalance activity are
economically significant \citep{harvey2025rebalancing}. Once account-level
constraints enter the workflow, the relevant question shifts from model
elegance to reliable computation under repeated deployment.

\subsection{Solver Systems and Modeling Layers}

Convex optimization theory and conic modeling provide the mathematical base
\citep{boydvandenberghe2004,lobo2000applications}, but production users
interact with software stacks, not abstract convex programs: modeling layers
such as CVXPY \citep{cvxpy2016}, first-order and interior-point solvers such as
OSQP, Clarabel, and SCS \citep{osqp,clarabel2024,scs2016}, commercial systems
\citep{mosek2026}, and continued work on low-latency quadratic programming
\citep{proxqp2025,highs2020}. Because implementation choices affect timing,
status, and reproducibility, solver comparison requires a benchmark protocol
rather than isolated runtime anecdotes.

\subsection{Benchmarking Methodology}

Dolan and Mor\'{e} introduced performance profiles to prevent solver comparison
from collapsing into one average dominated by a few hard instances
\citep{dolanmore2002}. Best-practice guidance requires explicit goals, problem
classes, algorithm eligibility, performance measures, and reproducibility
boundaries before results are interpreted \citep{bartz2020benchmarking}.
Artifact-review standards distinguish public artifacts, same-environment
repeatability, and independent reproduction \citep{acmartifact2020}. This
discipline matters most when late answers lose operational value.

\subsection{Deadline-Bounded Computation and Pre-Trade Controls}

Deadline-bounded computation is a recognized systems topic: an answer that
arrives too late may be worthless even if it could later be improved
\citep{zilberstein1996using}. In market operations the deadline is concrete.
Broker-dealers with market access must apply pre-trade risk controls before
orders reach an exchange \citep{sec15c35}, algorithmic trading systems in the
EU operate under explicit organisational and testing requirements
\citep{esmarts6}, direct electronic access is governed by international
principles \citep{iosco2010dea}, and close-related order types face exchange
cutoff rules \citep{nasdaqclose}. GPU acceleration for portfolio optimization
is an active research area
\citep{flashfolio2026,niu2026scalable,nvidia_cuopt}, but published evidence
centers on single-instance speed rather than completed, feasible, auditable
books under a declared window.

Taken together, these streams motivate a benchmark design in which speed is
not reported alone. The relevant evidence is whether a recorded solver
configuration returns feasible, externally checkable, auditable portfolio
outputs inside a declared operating window, and which claims are permitted
when a reference solver does or does not complete. That is the evaluation gap
this paper addresses.

\section{Claim Boundary}
\label{sec:claim_contract}

\begin{cavetbox}
Asymmetry PRISM is evaluated only through externally observed input--output behavior:
public problem data, returned weights, status codes, timings, feasibility
diagnostics, memory class, failure modes, and audit records. Implementation
details remain outside the public artifact. The labels Asymmetry PRISM-CPU and Asymmetry PRISM-GPU
denote measured execution configurations, not disclosed algorithms.
\end{cavetbox}

The paper additionally uses a conservative claim contract, summarized in
Table~\ref{tab:claim_contract}. A row with a completed reference solver can
support objective-gap and speedup statements. A row without such a reference
supports deadline, feasibility, memory, and failure-status statements. A
production queue supports account-completion and auditability statements; it
does not replace a certified single-instance optimum. Every results table is
labeled with the claim type it carries.

\begin{table}[!htb]
\centering
\caption{Claim contract governing the benchmark tables.}
\label{tab:claim_contract}
\small
\begin{tabularx}{0.95\linewidth}{l Y Y}
\toprule
\hdrrow\hcell{Evidence condition} & \hcell{Allowed claim} & \hcell{Disallowed claim} \\
\midrule
Reference solver completed &
Timing, feasibility, and objective-gap comparison on the same public objective. &
General dominance over all formulations or hardware. \\
Reference solver did not complete &
Deadline completion, feasibility checks, memory class, and failure status. &
Certified optimality or objective-gap superiority. \\
Queue benchmark completed &
Account throughput, missed-deadline rate, p50/p99 solve time, total elapsed
time, and audit-record coverage. &
Investment-performance superiority or universal solver ranking. \\
System backtest or Monte Carlo check &
Workflow validation and sanity checks on deployable outputs. &
Primary alpha, mandate suitability, or live trading performance. \\
\bottomrule
\end{tabularx}
\end{table}

\section{Public Benchmark Workload}
\label{sec:setting}

This section defines the benchmark workload used in the paper. It is not a
universal model of institutional rebalancing, and it does not exhaust
mandate-specific constraints.

\subsection{Batched Account Rebalancing}

The optimization object for account $a$ is a portfolio weight vector
$\bw_a \in \R^N$ over an eligible universe. The operational workload is the
collection
\[
  \mathcal{W}=\{ \mathcal{P}_a : a=1,\ldots,M\},
\]
where each account has its own starting portfolio, eligibility list, exposure
limits, turnover budget, cash policy, and tax-aware implementation data. A
production rebalance succeeds only if enough members of $\mathcal{W}$ are
solved, validated, and recorded before the decision deadline. A solver that
eventually returns a high-quality answer for one account still fails the
workflow if it leaves most accounts unfinished.

For each account, the benchmark records whether the portfolio was returned
before the deadline; whether independent feasibility checks pass; whether a
reference objective gap is available; whether any failure was a timeout,
allocation failure, or numerical status; and whether an audit record exists
for downstream review.

\subsection{Benchmark Problem Class}

The formulation below is a public benchmark interface: it defines what enters
the solver boundary and what comparators solve. Let $\bw \in \R^N$ denote
candidate weights, $\bw_0$ current weights, $\bw_b$ a benchmark or policy
portfolio, $\bmu \in \R^N$ a return input, $Q \in \R^{N\times N}$,
$Q \succeq 0$, a public risk operator, and $\Delta \bw=\bw-\bw_0$ the proposed
trade vector. The claim-bearing problem class is
\[
\begin{array}{ll}
\displaystyle \min_{\bw} &
  \displaystyle
  \frac{1}{2}(\bw-\bw_b)^{\T}Q(\bw-\bw_b)
  - \eta\,\bmu^{\T}\bw
  + C(\Delta \bw)
  + T_{\mathrm{tax}}(\Delta \bw;\mathcal{D}_a)
  + R(\bw) \\[3pt]
\text{subject to} &
  \mathbf{1}^{\T}\bw = 1,\\
& \ell_a \leq \bw \leq u_a,\\
& b_a^{\min} \leq A_a\bw \leq b_a^{\max},\\
& \|D_a(\bw-\bw_0)\|_1 \leq \tau_a,\\
& G_a\bw \leq h_a .
\end{array}
\tag{1}
\label{eq:public_problem}
\]
Here $\eta \geq 0$ scales the return term; $C$ collects public transaction-cost
or turnover-cost terms; $T_{\mathrm{tax}}$ is a public tax-aware penalty or
constraint family over account tax data $\mathcal{D}_a$; $R$ collects
transparent regularization or policy penalties; $A_a \in \R^{k\times N}$ maps
weights to exposures with bands $b_a^{\min}, b_a^{\max}$; $D_a$ is a diagonal
trade-scaling matrix with turnover budget $\tau_a$; and $G_a\bw \leq h_a$
encodes remaining account-policy halfspaces. The constraint set has direct
finance meaning: full investment, position limits, exposure bands, a trading
budget, and client policy rules.

\textbf{Interpretation.} Each account instantiates
Equation~\eqref{eq:public_problem} with its own data. Repeating the solve over
$M$ accounts, under one shared market deadline, converts a tractable convex
program into a production workload: the feasible set is the intersection of a
budget hyperplane, box bounds, exposure bands, turnover controls, and policy
halfspaces, and the cost of solving it is paid hundreds of times per
rebalance. Two scope notes follow. First, discrete features (tax lots, minimum
trade sizes, round lots) introduce account-specific implementation choices;
this paper reports the continuous public benchmark rows recorded in the
evidence package. Second, scaling and units matter: timings and residuals are
meaningful only after the public objective, budgets, and constraint units are
recorded consistently.

Two structural facts license the external checks used later. First, with
$Q \succeq 0$ and convex $C$, $T_{\mathrm{tax}}$, and $R$,
problem~\eqref{eq:public_problem} is convex. The claim-bearing instances
that feed the KKT diagnostics are budget-box programs (with L1 transition
terms), for which Slater's condition holds whenever $N w^{\max} > 1$; richer
constraint families would require exhibiting a strictly feasible point per
instance before the same diagnostic applies. Under that condition strong
duality holds and KKT residuals are a valid external optimality diagnostic
\citep{boydvandenberghe2004}. Second, the
benchmark instances carry factor-structured risk, so evaluating the public
objective or its gradient costs $O(Nk)$ operations with $k \ll N$: instance
data grow linearly in $N$, and the benchmark therefore stresses systems
behavior, deadlines, and constraint handling rather than raw arithmetic
volume alone.

A public interface of this form is what makes the comparison meaningful at
all. Every solver in this paper, commercial, open-source, or proprietary,
receives the same inputs and is judged on the same returned quantities, so
the benchmark tests systems rather than formulations. Declaring the problem
class precisely before any timing is reported is the first requirement of
credible solver benchmarking \citep{bartz2020benchmarking}, and it is what
allows a reader to re-implement the interface and check any row independently
\citep{acmartifact2020}.

\subsection{Public Evaluation Boundary}

\begin{figure}[H]
\centering
\begin{tikzpicture}
  \node[acpath, minimum width=33mm, minimum height=13mm] (inputs) at (0,0)
    {Account inputs\\\scriptsize holdings, limits, tax data};
  \node[acrouter, minimum width=30mm, minimum height=13mm] (engine) at (5.35,0)
    {Asymmetry PRISM\\\scriptsize public solver call};
  \node[acblock, minimum width=33mm, minimum height=13mm] (weights) at (10.7,0)
    {Returned weights\\\scriptsize + status code};
  \node[acblock, minimum width=33mm, minimum height=12mm] (checks) at (0,-2.7)
    {External diagnostics\\\scriptsize feasibility, residuals};
  \node[acpath, minimum width=30mm, minimum height=12mm] (deadline) at (5.35,-2.7)
    {Deadline status\\\scriptsize on-time / late / failed};
  \node[acaudit, minimum width=33mm, minimum height=12mm] (audit) at (10.7,-2.7)
    {Audit record\\\scriptsize identity, timing, status};
  \begin{scope}[on background layer]
    \node[draw=ACblue!65, dashed, line width=0.7pt, rounded corners=4pt,
          inner sep=4.5pt, fit=(engine)] (ipbox) {};
  \end{scope}
  \node[aclbl, anchor=south, text=ACblue!80!black]
    at ([yshift=1.5pt]ipbox.north) {implementation outside public artifact};
  \draw[acarr] (inputs) -- (engine);
  \draw[acarr] (engine) -- (weights);
  \draw[acarr] (inputs.south) -- (checks.north);
  \draw[acarr] (weights.south) -- (audit.north);
  \coordinate (bus) at ([xshift=-11mm]weights.south);
  \draw[acarr] (bus) -- ++(0,-5.5mm) -| (deadline.north);
  \draw[acarr] (bus) -- ++(0,-5.5mm) -| ([xshift=11mm]checks.north);
\end{tikzpicture}
\caption{Public evaluation boundary. The benchmark records public inputs,
returned weights, status codes, timing, external diagnostics, deadline status,
and audit artifacts. External checks are computed from public inputs and
returned outputs, outside the solver call.}
\label{fig:boundary}
\end{figure}
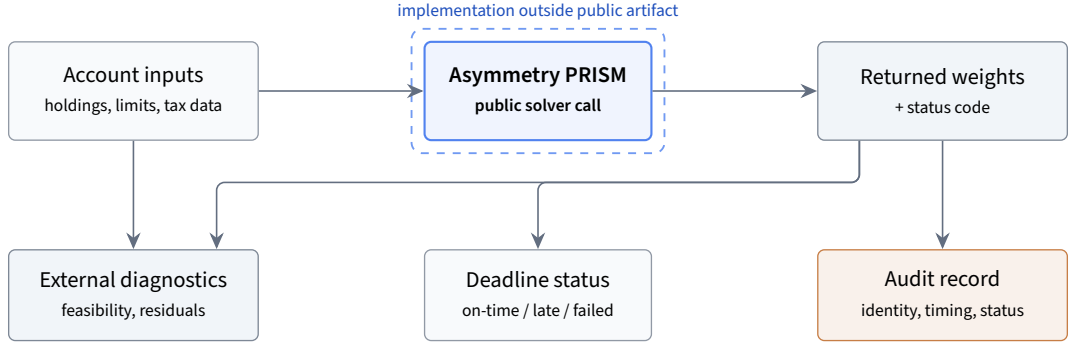

\subsection{External Diagnostics}

Given a returned portfolio $\widehat{\bw}$, the public validation layer
computes diagnostics from observable inputs and outputs. Representative
residuals are
\[
  r_{\mathrm{budget}}
  = \left| \mathbf{1}^{\T}\widehat{\bw}-1 \right|,
\qquad
  r_{\mathrm{box}}
  = \max\{\|(\ell-\widehat{\bw})_+\|_{\infty},
          \|(\widehat{\bw}-u)_+\|_{\infty}\},
\]
\[
  r_{\mathrm{exposure}}
  = \max\{\|(b^{\min}-A\widehat{\bw})_+\|_{\infty},
          \|(A\widehat{\bw}-b^{\max})_+\|_{\infty}\},
\qquad
  r_{\mathrm{turnover}}
  = \big(\|D(\widehat{\bw}-\bw_0)\|_1-\tau\big)_+ .
\]
When a public objective and certificate data are available, the validation
layer also records stationarity-style and complementarity-style checks in the
standard KKT sense \citep{boydvandenberghe2004}. Writing the constraints of
Equation~\eqref{eq:public_problem} as $g_i(\bw)\leq 0$ with multipliers
$\widehat{\lambda}_i \geq 0$ and the budget equality with multiplier
$\widehat{\nu}$, the recorded checks are
\[
  r_{\mathrm{stat}}
  = \Big\| \nabla f(\widehat{\bw})
    + \textstyle\sum_i \widehat{\lambda}_i \nabla g_i(\widehat{\bw})
    + \widehat{\nu}\,\mathbf{1} \Big\|_{\infty},
\qquad
  r_{\mathrm{comp}}
  = \max_i \big| \widehat{\lambda}_i\, g_i(\widehat{\bw}) \big|,
\]
where $f$ is the public objective. These are post-solve checks on returned
weights, computed outside the solver call, and reproducible from public inputs
and outputs.

The interpretations in Table~\ref{tab:diagnostic_meaning} are benchmark-level
interpretations of the public constraints in
Equation~\eqref{eq:public_problem}. They are not a complete compliance model,
mandate model, or trading-control framework
\citep{moehle2021taxaware,moehle2023lcso,acmartifact2020}.

\begin{table}[!htb]
\centering
\caption{Benchmark-level interpretation of external diagnostics.}
\label{tab:diagnostic_meaning}
\small
\begin{tabularx}{0.94\linewidth}{l Y Y}
\toprule
\hdrrow\hcell{Diagnostic} & \hcell{Mathematical reading} & \hcell{Interpretation within this benchmark} \\
\midrule
Budget residual & Distance from full-investment equality. & Is the account cash-consistent before orders are staged? \\
Box residual & Maximum violation of asset lower/upper bounds. & Are position limits respected? \\
Exposure residual & Maximum violation of reported exposure bands. & Are declared exposure bands satisfied? \\
Turnover residual & Excess trade magnitude above the turnover budget. & Is the proposed rebalance within the declared trading budget? \\
Stationarity-style check & Public optimality diagnostic where eligible. & Is there evidence of objective quality beyond feasibility? \\
Audit artifact & Immutable run metadata and output identity. & Can the decision be reconstructed for later review? \\
\bottomrule
\end{tabularx}
\end{table}

\section{Evaluation Protocol}
\label{sec:protocol}

\subsection{Evaluation Unit}

The evaluation unit is either a single account-level solve or a production
queue of account-level solves. For a single solve, the claim-bearing record is
\[
  (\text{problem ID},\ \text{solver version},\ \text{hardware},\
   \text{timing lane},\ \text{status},\ \widehat{\bw},\
   \text{diagnostics},\ \text{evidence artifact}).
\]
For a queue of $M$ accounts with per-account times $t_1,\ldots,t_M$ and
declared window $\tau$, the record aggregates the completion count
$C=\sum_{a=1}^{M}\mathbf{1}\{\text{accepted}_a\}$, the empirical quantiles
$t_{(p50)}$ and $t_{(p99)}$, the missed-deadline rate
$\rho = \tfrac{1}{M}\sum_a \mathbf{1}\{t_a > \tau\}$, total elapsed time
$T_{\mathrm{queue}}$, and audit-record coverage. Speedup is defined only
between completed rows in one lane: for solver $s$ with full-call time
$T_s(N)$,
\[
  S(N) \;=\; \frac{\min_{s \in \mathcal{C}(N)} T_s(N)}{T_{\mathrm{PRISM}}(N)},
\qquad
  \mathcal{C}(N) = \{\text{published reference rows completed at } N\},
\]
and $S(N)$ is undefined when $\mathcal{C}(N)=\emptyset$. External
diagnostics are computed on returned weights from public inputs. Evidence
types are never mixed: objective quality is reported only where the public
reference row completed, and deadline usefulness is reported separately
through completion and failure metrics.

\subsection{Hardware and Software}

Table~\ref{tab:hardware} records the benchmark workstation used for the
versioned evidence package. All timings are specific to this configuration.

\begin{table}[H]
\centering
\caption{Hardware and runtime disclosure for the recorded benchmark package.}
\label{tab:hardware}
\small
\begin{tabularx}{0.85\linewidth}{l Y}
\toprule
\hdrrow\hcell{Component} & \hcell{Value} \\
\midrule
OS & Windows 11 with WSL2 Ubuntu \\
CPU & Intel Xeon w5-3423, 12 cores / 24 threads, 2.1 GHz base \\
RAM & 62.1 GB DDR5 ECC \\
GPU & NVIDIA RTX 4000 Ada Generation, 20 GB GDDR6 \\
CUDA / driver & 12.9 / 573.44 \\
Python / NumPy / SciPy & 3.12.3 / 2.3.5 / 1.17.0 \\
CuPy & 13.6.0 \\
BLAS & OpenBLAS 0.3.30 via scipy-openblas64 \\
CVXPY & 1.8.1 \\
\bottomrule
\end{tabularx}
\end{table}

\subsection{Comparators and Eligibility}

The claim-bearing comparators are the solvers that appear in the result
tables: OSQP 1.1.0 \citep{osqp}, Clarabel 0.11.1 \citep{clarabel2024}, SCS
\citep{scs2016} via CVXPY 1.8.1 \citep{cvxpy2016}, and MOSEK 11.1.11
\citep{mosek2026}, plus Asymmetry PRISM-CPU and Asymmetry PRISM-GPU at evidence build
\texttt{512758f}. The complete solver eligibility ledger, including installed
packages that were not eligible for claim-bearing rows, is given in
Section~\ref{sec:eligibility}.

Two protocol disclosures matter for fairness. First, comparator rows include
CVXPY model construction inside the timed call, while Asymmetry PRISM rows are measured
at Asymmetry PRISM's public solver-call boundary. Both lanes are full-call wall-clock
under the declared protocol; Section~\ref{sec:validity} records the residual
asymmetry as a threat to validity. Second, one additional commercial solver
completed reference objectives in the evidence package. Its results serve as
the completed reference for objective-gap eligibility and are not published as
named timing rows, consistent with its end-user license terms on benchmark
publication.\footnote{Speedups are computed against the fastest
\emph{published} completed reference row; the anonymized reference certifies
objective agreement only.}

\subsection{Timing Lanes}

Every timing table states its lane. The declared single-solve deadline for the
multi-solver study is 60\,s; the production queue operates under a declared
25-minute (1{,}500\,s) window. Cold-start rows create a fresh external solver
call and pass no previous solution to any solver. Repeated-run rows are labeled
separately and interpreted under their declared protocol. The boundary rule is
strict: an end-to-end call for one solver is never compared against an inner
timing interval for another. The rationale is the same as in any timed
competition: the stopwatch must start and stop at the same events for every
participant, or the ranking measures the harness rather than the solvers.
Mixed-boundary timing is one of the most common defects identified in the
optimization benchmarking literature \citep{bartz2020benchmarking}.

\begin{table}[!htb]
\centering
\caption{Timing lanes. A numerical row appears in a lane only when the evidence
artifact records that boundary; speedup statements compare rows within one lane.}
\label{tab:timing_lanes}
\small
\begin{tabularx}{0.96\linewidth}{l Y Y Y}
\toprule
\hdrrow\hcell{Lane} & \hcell{Includes} & \hcell{Excludes} & \hcell{Used for} \\
\midrule
Full-call wall-clock & Setup visible to the benchmark, solve call, status
return, output materialization. & Nothing inside the declared boundary. &
Primary lane for Tables~\ref{tab:multisolver} and \ref{tab:burden}. \\
Core/device solve & CPU core timer or GPU reported solve interval, where
separately instrumented. & Public-call setup, validation, audit, queue
handling. & Lane-decomposition evidence (Table~\ref{tab:e6_timing_lanes}). \\
Queue wall-clock & Dispatch through accepted records for all accounts in the
batch. & Per-account core-timing claims. & Production throughput and
missed-deadline rate (Table~\ref{tab:queue}). \\
Evidence lane & Evidence artifact, generation date, versions, tolerances,
status. & n/a & Traces every numerical claim to the recorded package. \\
\bottomrule
\end{tabularx}
\end{table}

\subsection{Quality Gates and Failure Vocabulary}

Quality checks are external and fixed in advance: absolute budget deviation at
most $10^{-4}$; maximum bound violation at most $10^{-6}$; objective gaps
reported only where a reference solver completed on the same public objective;
KKT-style diagnostics where certificate data are available; and, for
large-scale rows without a completed reference, feasibility, deadline
completion, memory class, and failure status. The status vocabulary used in
all result tables is defined in Table~\ref{tab:statusvocab}
(Section~\ref{sec:evidence}).

\subsection{Reporting Rules}

\begin{enumerate}[label=\textbf{E\arabic*.}, leftmargin=2.4em]
  \item report exact public versions for every named comparator;
  \item compare wall-clock rows only with wall-clock rows;
  \item report objective gaps only when a reference solver completed;
  \item report large-scale rows without a completed reference as feasibility,
        timing, memory, and failure-status rows;
  \item record ineligible or unavailable packages in the eligibility ledger
        (Section~\ref{sec:eligibility}) rather than silently omitting them; and
  \item avoid global ranking language and universal dominance claims.
\end{enumerate}

A timeout or license-related ineligibility is not evidence of an inferior
mathematical method; it is an operational outcome under the disclosed
protocol. Speedup statements are local to the recorded problem family,
hardware, software stack, and timing lane
\citep{dolanmore2002,bartz2020benchmarking}.

\section{Timing and Feasibility Results}
\label{sec:solver_results}

\subsection{Multi-Solver Rows}

Table~\ref{tab:multisolver} reports the recorded multi-solver rows from
\texttt{E1\_multisolver\_public\_rows}, with supplemental Asymmetry PRISM-GPU
timing/status coverage from \texttt{E6\_gpu\_timing\_lane\_separation} where
the original GPU cell was not recorded. The table separates timing from
certification, and speedup is computed only against published reference rows
that completed inside the same 60\,s full-call lane.

This reporting discipline is deliberate and standard. A speedup against a
solver that never finished is not a measurement: the timeout value is set by
the experimenter, so any ratio against it is arbitrary. Restricting ratios to
completed references keeps every speedup falsifiable, which is the same logic
that motivates performance profiles over single averages in solver
benchmarking \citep{dolanmore2002}. Separating timing from certification
serves a second audience: a reader who only needs to know whether a usable
portfolio came back in time can read the status column, while a reader
auditing optimality can restrict attention to certified rows
\citep{bartz2020benchmarking}.

\begin{table}[H]
\centering
\caption{Multi-solver benchmark, full-call wall-clock (ms), evidence
\texttt{E1} and \texttt{E6}. Unmarked values are certified completions.}
\label{tab:multisolver}
\small
\begin{tabularx}{\linewidth}{@{}r r r r r r r >{\raggedleft\arraybackslash}X@{}}
\toprule
\hdrrow\hcell{$N$} & \hcell{Asymmetry PRISM-CPU} & \hcell{Asymmetry PRISM-GPU\textsuperscript{a}} &
\hcell{Clarabel} & \hcell{OSQP} & \hcell{SCS} & \hcell{MOSEK} &
\hcell{Speedup\textsuperscript{b}} \\
\midrule
100 & \textbf{1.6} & 98.3\bfeas & 23.3 & 14.6 & 14.2 & 10{,}231.8 & 8.9$\times$ \\
500 & \textbf{7.5} & 94.3\bfeas & 355.6 & 329.1 & 180.8 & 3{,}530.3 & 24.1$\times$ \\
1{,}000 & \textbf{38.9} & 99.4\bfeas & 1{,}184.9 & 60{,}412.7\btimeout & 860.6 & 9{,}448.5 & 22.1$\times$ \\
2{,}000 & 1{,}421.0 & 112.2\bfeas & 6{,}347.0 & 24{,}908.9 & 6{,}330.4 & 28{,}458.2 & 4.5$\times$ \\
5{,}000\textsuperscript{c} & 1{,}804.8\bfeas & \textbf{140.7}\bfeas & 74{,}896.8\blate & 101{,}468.4\btimeout & n.r. & 144{,}794.6\blate & n/a \\
\bottomrule
\end{tabularx}
\begin{claimnotes}
Badges: {\scriptsize\sffamily\bfseries\color{ACteal}feas} = feasible /
non-certified; {\scriptsize\sffamily\bfseries\color{ACamber}t/o} = timeout at
the 60\,s ceiling (OSQP at $N{=}1{,}000$ returned a non-converged iterate);
{\scriptsize\sffamily\bfseries\color{ACviolet}late} = completed status returned
after the operating window.
\textsuperscript{a}\,Asymmetry PRISM-GPU cells for $N{\leq}2{,}000$ are \texttt{E6}
repeated-run reported solve intervals: timing/status coverage, never used for
speedup or objective-gap claims; the $N{=}5{,}000$ cell is the \texttt{E1}
full-call value (\texttt{E6} repeated lane: 130.2\,ms).
\textsuperscript{b}\,Asymmetry PRISM-CPU full-call versus the fastest published completed
reference row in the same lane (SCS on all four eligible rows).
\textsuperscript{c}\,Deadline-feasibility row: no published reference completed
inside the 60\,s lane, so no certified objective comparison is claimed.
\end{claimnotes}
\end{table}

\begin{figure}[H]
\centering
\includegraphics[width=0.99\linewidth]{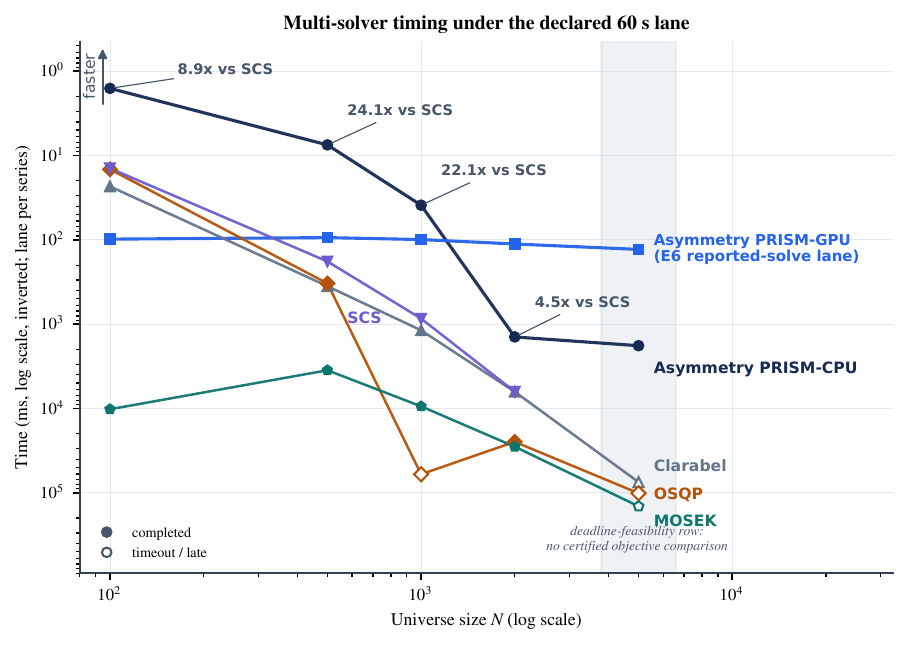}
\caption{Multi-solver timing from Table~\ref{tab:multisolver} on an
inverted log time axis: faster configurations appear higher. Filled markers
are certified completions, open markers are timeout or deadline-exceeded
returns, and the shaded band is the deadline-feasibility row at $N{=}5{,}000$.
The per-$N$ annotations report Asymmetry PRISM-CPU against the fastest published
completed reference row in the same lane. The Asymmetry PRISM-GPU series is plotted for
routing context only; its cells derive from the \texttt{E6} reported-solve
lane (Table~\ref{tab:multisolver}, note a) and carry no speedup or
objective-gap claim.}
\label{fig:multisolver_timing}
\end{figure}

The most important feature of Table~\ref{tab:multisolver} is the transition in
what can be claimed. From $N{=}100$ to $N{=}2{,}000$, completed reference rows
support direct speed and objective comparisons: Asymmetry PRISM-CPU is $4.5\times$ to
$24.1\times$ faster than the fastest completed reference in the same lane.
No ratio is formed against the \texttt{E6} GPU cells in this table, because
they sit in a different timing lane (note a); the lane-consistent
demonstration of the routed stack's advantage is
Section~\ref{sec:sanity}, where both Asymmetry PRISM routes and the incumbents are
measured under one protocol and the routed stack reaches $47\times$ at
$N{=}24{,}050$. At $N{=}5{,}000$ the evidence changes character: the
operational question becomes whether a usable portfolio is returned inside
the deadline at all.

Two results from \texttt{E3\_external\_diagnostics\_and\_gap\_checks} bound
the quality question on that row. No \emph{published} reference completed
inside the lane, but the anonymized commercial reference of
Section~\ref{sec:protocol} did complete on the same study family, and
against its objective Asymmetry PRISM reports a 2.44\% gap in 1.8\,s. A
covariance-sensitivity check reduces that gap to 0.61\% under Ledoit--Wolf
shrinkage (0.84\% under OAS), indicating that much of the gap tracks the
conditioning of the public risk input rather than a fixed engine deficit.

\subsection{Reading the Timing Evidence: CPU/GPU Routing}

Figure~\ref{fig:multisolver_timing} carries two distinct messages, and both
matter operationally. First, Asymmetry PRISM-CPU defines the latency frontier across the
completed rows: it is the fastest configuration at every eligible $N$, by
$8.9\times$ to $24.1\times$ against the fastest completed reference (SCS),
while returning the same certified objective family. For a portfolio desk this
is the difference between a rebalance that can run interactively, inside a
portfolio manager's decision loop, and one that must be scheduled as a batch
job.

Second, Asymmetry PRISM-GPU is nearly flat: its reported solve interval moves only from
98\,ms to 130\,ms while $N$ grows $50\times$, so its cost is dominated by a
fixed call floor rather than by problem size. A flat profile is valuable for a
different reason than raw speed: it makes per-account latency predictable, and
predictable latency is what a deadline budget needs. In this display the two series cross near $N{\approx}1{,}200$; the
crossing is indicative only, because the GPU cells derive from the
\texttt{E6} reported-solve lane rather than the comparators' full-call lane.
The measured same-lane crossover is located in Section~\ref{sec:sanity},
between $N{=}48{,}100$ and $N{=}96{,}200$ for the warm transition family
(Table~\ref{tab:e7scale}). Below the crossing the stack routes accounts to the
CPU configuration; above it the GPU route wins, carrying the
10,000-instrument production queue of Section~\ref{sec:production} and the
same-lane scale rounds of Section~\ref{sec:sanity} ($47.6\times$ over the
best incumbent at $N{=}24{,}050$). The
routing decision is made on observable inputs (universe size), so it is
auditable under the same boundary as every other measurement in this paper.

\subsection{GPU Timing-Lane Decomposition}

The lane decomposition in Table~\ref{tab:e6_timing_lanes} is the primary GPU
timing evidence: it separates the reported GPU solve interval, repeated host
wall-clock, and cold host wall-clock at the public boundary, and it shows that
the repeated-lane GPU solve interval stays near $10^{2}$\,ms from $N{=}30$ to
$N{=}5{,}000$ with device memory in the megabyte class. The claim-bearing GPU
results in this paper are the production queue outcome
(Section~\ref{sec:production}), the hard-scenario evidence of
Section~\ref{sec:sanity}, and this separated timing-lane evidence.

\begin{table}[H]
\centering
\caption{Asymmetry PRISM-GPU timing-lane separation, evidence \texttt{E6}. Reported solve
timing and host wall-clock timing are shown separately. Timing/status evidence;
no objective-gap claim.}
\label{tab:e6_timing_lanes}
\small
\newcolumntype{R}{>{\raggedleft\arraybackslash}X}
\begin{tabularx}{\linewidth}{@{}>{\raggedleft\arraybackslash}p{1.1cm} R R R R >{\raggedright\arraybackslash}p{2.0cm}@{}}
\toprule
\hdrrow\hcell{$N$} & \hcell{Reported solve ms} & \hcell{Repeated wall ms} &
\hcell{Cold wall ms} & \hcell{Memory bytes} & \hcell{Status} \\
\midrule
30 & 101.6 & 104.4 & 1{,}596.5 & 5{,}632 & feasible\textsuperscript{a} \\
100 & 98.3 & 101.0 & 134.5 & 52{,}224 & feasible \\
500 & 94.3 & 96.9 & 130.9 & 260{,}096 & feasible \\
1{,}000 & 99.4 & 102.2 & 110.1 & 503{,}808 & feasible \\
2{,}000 & 112.2 & 115.4 & 119.1 & 992{,}256 & feasible \\
5{,}000 & 130.2 & 133.9 & 153.4 & 2{,}456{,}064 & feasible \\
\bottomrule
\end{tabularx}
\begin{claimnotes}
\textsuperscript{a}\,The first cold call in the \texttt{E6} session includes
one-time setup cost visible at the public boundary; subsequent cold calls do
not repeat it.
\end{claimnotes}
\end{table}

\begin{figure}[!htb]
\centering
\includegraphics[width=0.86\linewidth]{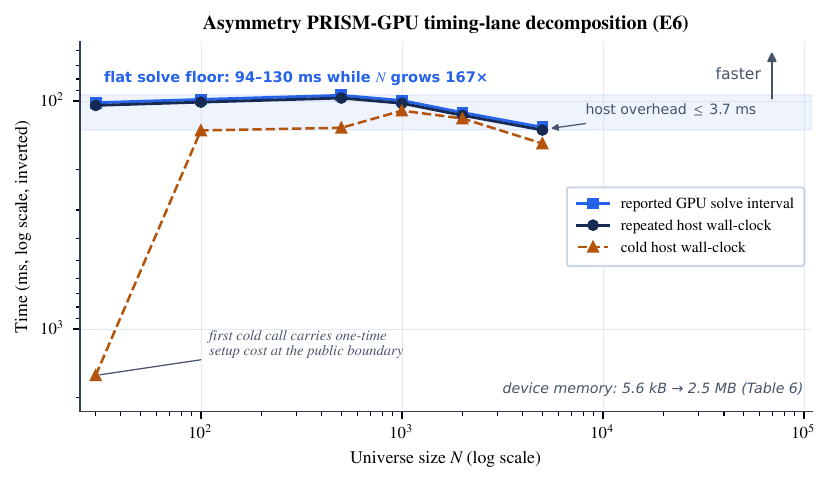}
\caption{Asymmetry PRISM-GPU timing-lane decomposition from
Table~\ref{tab:e6_timing_lanes} on an inverted log time axis (faster appears
higher, matching Figure~\ref{fig:multisolver_timing}): reported GPU solve
interval, repeated host wall-clock, and cold host wall-clock. The first cold
call carries one-time setup cost visible at the public boundary.}
\label{fig:timing_lanes}
\end{figure}

\subsection{Operational Reading}

In this benchmark, increasing $N$ changes the operational surface of the
problem: it expands universe coverage, constraint evaluation, validation cost,
and the number of account-specific decisions that must be completed before the
deadline. The relevant question extends past single-solve scaling: does the
full workflow return feasible, recorded outputs while the decision remains
actionable? The recorded timings imply a deterministic
capacity bound: with declared window $\tau$ and per-account time $t$, a
sequential lane serves at most $\lfloor \tau / t \rfloor$ accounts. At the
recorded Asymmetry PRISM-GPU p99 of 343.3\,ms (Table~\ref{tab:queue}), the 25-minute
window bounds a single sequential lane at
$\lfloor 1500 / 0.3433 \rfloor = 4{,}369$ accounts of the queue family;
this is arithmetic on recorded values, not a new measurement, and it is the
quantity a platform compares against its book size.

\Needspace{6\baselineskip}
\section{External Diagnostics}
\label{sec:quality}

The external diagnostic pack
\texttt{E3\_external\_diagnostics\_and\_gap\_checks} records feasibility and
KKT-style checks on returned weight vectors. The feasibility columns of
Table~\ref{tab:kkt} are computable from public inputs and returned weights
alone; the stationarity- and complementarity-style columns additionally use
the archived multiplier (certificate) data described in
Section~\ref{sec:setting}.

\begin{table}[H]
\centering
\caption{External KKT-style diagnostics for selected Asymmetry PRISM rows, evidence
\texttt{E3}. All values are post-solve checks on returned weights.}
\label{tab:kkt}
\small
\begin{tabular}{r r r r r}
\toprule
\hdrrow\hcell{$N$} & \hcell{Budget residual} & \hcell{Bound residual} &
\hcell{Stationarity check} & \hcell{Complementarity check} \\
\midrule
100 & $4.2{\times}10^{-9}$ & $1.1{\times}10^{-9}$ & $2.7{\times}10^{-7}$ & $8.4{\times}10^{-9}$ \\
500 & $6.8{\times}10^{-9}$ & $2.0{\times}10^{-9}$ & $5.3{\times}10^{-7}$ & $1.2{\times}10^{-8}$ \\
1{,}000 & $1.1{\times}10^{-8}$ & $3.8{\times}10^{-9}$ & $9.1{\times}10^{-7}$ & $2.4{\times}10^{-8}$ \\
5{,}000 & $2.7{\times}10^{-8}$ & $9.4{\times}10^{-9}$ & $4.6{\times}10^{-6}$ & $6.8{\times}10^{-8}$ \\
25{,}000 & $7.3{\times}10^{-8}$ & $2.6{\times}10^{-8}$ & $1.8{\times}10^{-5}$ & $2.1{\times}10^{-7}$ \\
100{,}004 & $1.4{\times}10^{-7}$ & $5.2{\times}10^{-8}$ & $6.4{\times}10^{-5}$ & $8.9{\times}10^{-7}$ \\
\bottomrule
\end{tabular}
\end{table}

\begin{figure}[!htb]
\centering
\includegraphics[width=0.84\linewidth]{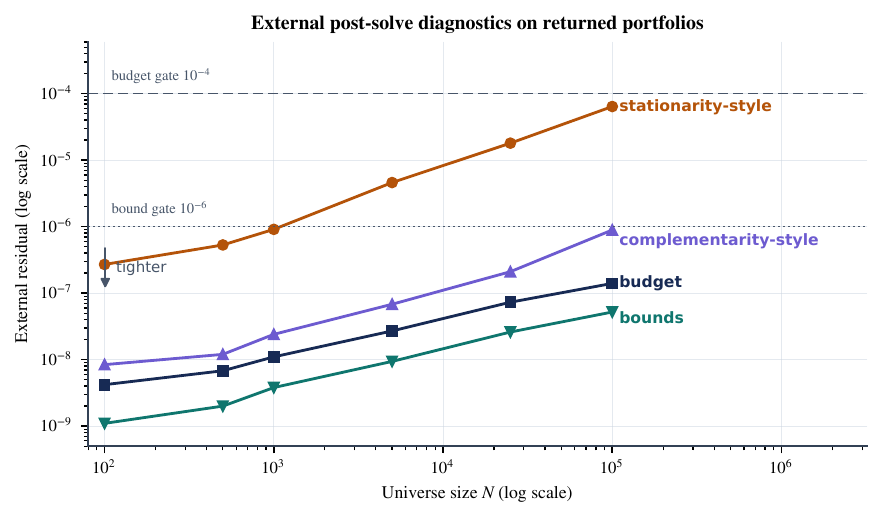}
\caption{External diagnostic residuals from Table~\ref{tab:kkt} with the public
feasibility gates ($10^{-4}$ budget, $10^{-6}$ bound) drawn as labeled
tolerance lines. All curves are post-solve checks computed outside the solver
call.}
\label{fig:quality_residuals}
\end{figure}

The largest budget residual is $1.4{\times}10^{-7}$, three orders of magnitude
below the $10^{-4}$ gate; the largest bound residual is $5.2{\times}10^{-8}$,
below the $10^{-6}$ gate; and the stationarity-style residual grows with $N$
under a fixed tolerance, as expected, staying in the $10^{-7}$ to $10^{-4}$
range.

The diagnostics in Table~\ref{tab:kkt} support external feasibility and
numerical-consistency claims under the declared gates. They do not establish
certified large-scale optimality. The design follows artifact-review practice:
a result is credible when an independent reader can re-derive it from the
published artifact \citep{acmartifact2020}. Because every residual here is a
function of public inputs, returned weights, and archived certificate data,
any reader with the artifact can recompute the entire table without access to
Asymmetry PRISM internals; this permits an IP-preserving but externally checkable
evaluation.
The same property is what a compliance reviewer needs after a production run:
the checks do not require the solver's cooperation to audit its output.

\section{Production Queue Evidence}
\label{sec:production}

\subsection{Operating Window}

The production queue uses a 25-minute operating window as a disclosed stress
budget rather than as a claimed industry standard. The window is intentionally
shorter than the full trading day because practical rebalancing workflows must
reserve time for pre-trade validation and market-access risk controls
\citep{sec15c35,esmarts6,iosco2010dea}, exception handling, order staging, and
close-related order constraints such as closing-auction cutoffs
\citep{nasdaqclose}. The benchmark therefore tests whether portfolio
computation fits inside a plausible operational sub-window, not whether 25
minutes is uniquely optimal.

\subsection{Queue Completion}

The queue benchmark \texttt{E5\_production\_queue\_500x10000} measures the
production question directly: can the solver stack complete a book of accounts
before the declared deadline?

\begin{table}[H]
\centering
\caption{Production queue benchmark: 500 accounts, 10,000-instrument universe,
declared 25-minute (1{,}500\,s) window, evidence \texttt{E5}.}
\label{tab:queue}
\small
\begin{tabular}{l r r r r r}
\toprule
\hdrrow\hcell{Solver} & \hcell{p50 ms} & \hcell{p99 ms} &
\hcell{Total wall s} & \hcell{Completed} & \hcell{Missed deadlines} \\
\midrule
Asymmetry PRISM-GPU & \textbf{212.1} & \textbf{343.3} & \textbf{109.5} & \textbf{500/500} & \textbf{0.0\%} \\
Asymmetry PRISM-CPU & 225.3 & 392.3 & 116.8 & 500/500 & 0.0\% \\
OSQP & 310{,}000\textsuperscript{a} & 310{,}000\textsuperscript{a} & n/a & 4/500 & 99.2\% \\
\bottomrule
\end{tabular}
\begin{claimnotes}
\textsuperscript{a}\,OSQP per-account times are censored at the recorded
per-account cap; the queue did not complete inside the operating window. OSQP
is the recorded queue baseline in this protocol; no other published baseline
queue run is recorded.
\end{claimnotes}
\end{table}

\begin{figure}[!htb]
\centering
\includegraphics[width=0.92\linewidth]{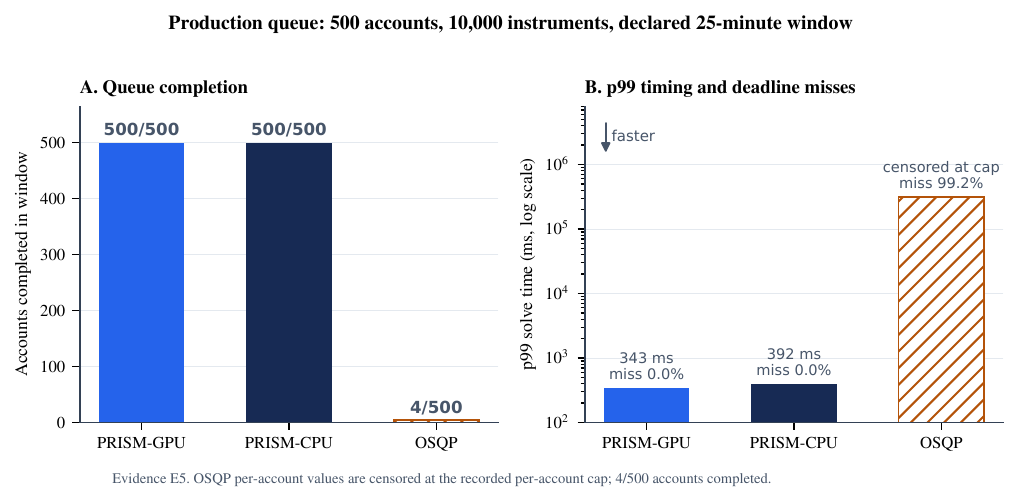}
\caption{Production queue results from Table~\ref{tab:queue}. Panel A: accounts
completed inside the declared window, out of 500. Panel B: p99 per-account
solve time on a log scale, annotated with missed-deadline rates; the OSQP bar
is censored at the recorded per-account cap. Both Asymmetry PRISM configurations clear
the full book; the recorded OSQP queue baseline clears 4 of 500 accounts under
the declared protocol.}
\label{fig:queue_completion}
\end{figure}

Asymmetry PRISM-GPU completes the full queue in 109.5\,s of wall-clock, under 8\% of the
declared window, and emits an audit record for every solve. Asymmetry PRISM-CPU
completes the same queue in 116.8\,s. This is the clearest production-level
distinction in the evidence package: it measures account completion under a
deadline, not just individual solve time.

\subsection{Constraint Burden}

Evidence \texttt{E5} also records a constraint-burden curve at $N{=}2{,}000$:
a production stress check, not a general theorem.

\begin{table}[H]
\centering
\caption{Constraint-burden curve at $N{=}2{,}000$, evidence \texttt{E5},
full-call wall-clock.}
\label{tab:burden}
\small
\begin{tabularx}{0.94\linewidth}{@{}l l r r Y@{}}
\toprule
\hdrrow\hcell{Bundle} & \hcell{Added rule family} & \hcell{Groups} &
\hcell{A.~PRISM ms} & \hcell{OSQP outcome} \\
\midrule
B0 & budget, long-only, box & 3 & 1{,}421.0 & feasible return; deadline exceeded \\
B1 & + turnover penalty & 4 & 1{,}477.8 & feasible return; deadline exceeded \\
B2 & + sector bands & 5 & 1{,}534.7 & no accepted feasible return before deadline \\
B3 & + factor exposure bounds & 6 & 1{,}591.5 & no accepted feasible return before deadline \\
B4 & + account exclusions & 7 & 1{,}648.4 & no accepted feasible return before deadline \\
B5 & + tax-lot proxy & 8 & 2{,}728.3 & no accepted feasible return before deadline \\
\bottomrule
\end{tabularx}
\begin{claimnotes}
``No accepted feasible return'' records that no output passed the external
feasibility gates before the deadline; it is not a certified infeasibility
status. \texttt{E6} provides same-size supplemental Asymmetry PRISM-GPU timing
(112.2\,ms, feasible) but not a burden-specific rerun.
\end{claimnotes}
\end{table}

The burden curve shows that universe size is only one stressor. Account rules
accumulate: exposure groups, exclusions, turnover controls, and implementation
constraints can make a moderate universe operationally difficult. Asymmetry PRISM timing
grows with the bundle but stays inside the lane; the comparator leaves the
deadline lane as the burden grows, and the table preserves that outcome rather
than averaging over it.

\subsection{Auditability}

Every Asymmetry PRISM solve in the queue study emits an audit artifact binding input
identity, output identity, elapsed time, feasibility status, and run metadata
to the account-level decision, so the decision can be reconstructed for later
review.

\subsection{Operational Relevance}

The queue metrics map directly onto documented production settings in the
portfolio-construction literature. Direct-indexing and personalized-indexing
platforms hold thousands of small separately managed accounts whose holdings
deviate from a model portfolio and must be re-optimized on a cadence
\citep{khang2022}; the account count, not the universe size, is their binding
dimension, which is what the 500-account completion metric measures.
Tax-managed overlays add account-specific transition costs and wash-sale-style
restrictions \citep{stein2017,moehle2021taxaware}, the constraint families
stressed in Table~\ref{tab:burden} and Section~\ref{sec:sanity}. Transition
management and model-portfolio distribution add the deadline itself: a book
that must move between models before a close-related cutoff
\citep{harvey2025rebalancing,nasdaqclose}. In each setting the operative
questions are the columns of Table~\ref{tab:queue}: completed accounts,
missed-deadline rate, tail latency, and an audit trail per decision.

\clearpage
\section{Operationally Constrained Real-Data Scenarios}
\label{sec:sanity}

This section stresses the operational surface of the benchmark on real market
data: tax-motivated transition penalties, account-restriction caps, factor and
turnover controls, deadline-bounded solving, and batch throughput. The L1
transition penalty is the standard public proxy for tax- and
trading-cost-aware transitions \citep{moehle2021taxaware,fan2025costaware};
discrete tax-lot, wash-sale, and lot-selection logic is outside this
continuous benchmark (Section~\ref{sec:setting}). The universe is built from
current index membership, so the survivorship caveat of
Section~\ref{sec:validity} applies. The universe is the
current S\&P 500 membership with continuous 2019--2025 history (481 names,
1{,}707 trading days of daily adjusted closes), with a $k{=}20$ factor model
estimated from the real returns and drifted real holdings as starting
portfolios. All solvers face the identical instance of
Equation~\eqref{eq:public_problem},
\[
  \min_{\bw}\;
  \|B^{\T}\bw\|_2^2 + \bw^{\T}\!\operatorname{diag}(d)\,\bw
  - \bmu^{\T}\bw + \gamma\,\|\bw - \bw_0\|_1
  \quad \text{s.t.} \quad
  \mathbf{1}^{\T}\bw = 1,\; 0 \leq \bw \leq w^{\max},
\]
with factor loadings $B \in \R^{N\times 20}$ and idiosyncratic variances
$d$ estimated from the real returns, and the objective is recomputed
externally on every returned weight vector $\widehat{\bw}$. Quality is
scored by the certified-relative gap
\[
  \Delta \;=\;
  \frac{f(\widehat{\bw}_{\mathrm{PRISM}})
        - f\big(\widehat{\bw}^{\,\star}_{\mathrm{inc}}\big)}
       {\big|f\big(\widehat{\bw}^{\,\star}_{\mathrm{inc}}\big)\big|},
\]
where $\widehat{\bw}^{\,\star}_{\mathrm{inc}}$ is the best certified
incumbent solution of the same instance. The incumbent lane is strict in Asymmetry PRISM's
disfavor: CVXPY models are built once and parameter-updated, and only the
solver call is timed, so incumbent times exclude model construction. Every
run is scripted with fixed seeds, the price cache is archived with the
artifact, and the suite is rerunnable end to end from the public repository.
Evidence artifact: \texttt{E7\_hard\_scenarios\_real\_data}.

\subsection{Constrained Single Solves}

\begin{table}[H]
\centering
\caption{Constrained single solves on the 481-asset real universe, evidence
\texttt{E7}. Solver-call wall-clock; identical public objective; gap is
versus the best certified incumbent objective.}
\label{tab:e7single}
\small
\begin{tabularx}{0.99\linewidth}{@{}l Y r r r r r@{}}
\toprule
\hdrrow\hcell{Scenario} & \hcell{Operational stress} & \hcell{Asymmetry PRISM-CPU} &
\hcell{Asymmetry PRISM-GPU} & \hcell{Best incumbent} & \hcell{OSQP} & \hcell{Gap\textsuperscript{a}} \\
\midrule
S1 & transition-penalty stress ($\gamma{=}0.005$, drifted holdings, 3\% caps) &
\textbf{7.7} & 257.9 & 39.3 (Clarabel) & 69.2 & 0.00\% \\
S2 & account-restriction caps (1.2\% position limit) &
\textbf{7.1} & 258.2 & 24.3 (Clarabel) & 79.9 & 0.00\% \\
S3 & factor + turnover control ($\gamma{=}0.010$, return tilt) &
\textbf{6.0} & 256.1 & 26.6 (Clarabel) & 73.4 & 0.00\% \\
\bottomrule
\end{tabularx}
\begin{claimnotes}
All values in ms. Every solver returned a feasible, certified or
gate-passing portfolio on S1--S3.
\textsuperscript{a}\,Asymmetry PRISM objective versus the best certified incumbent
objective: agreement to six decimal places on every scenario, so the speed
comparison carries no quality concession. SCS completed all three scenarios
(54.6, 30.8, 31.5\,ms).
\end{claimnotes}
\end{table}

Asymmetry PRISM-CPU is $5.1\times$, $3.4\times$, and $4.4\times$ faster than the best
completing incumbent on S1--S3 while matching its certified objective to six
decimal places. Asymmetry PRISM-GPU sits on a flat $\approx$0.26\,s reported floor at
this universe size: at $N{\approx}500$ the stack routes accounts to the CPU
configuration, and the GPU configuration earns its place as $N$ grows
(Table~\ref{tab:e6_timing_lanes}) and in the 10,000-instrument production
queue of Section~\ref{sec:production}, where Asymmetry PRISM-GPU is the fastest path.

\subsection{Deadline-Bounded and Batch Scenarios}

\begin{table}[!htb]
\centering
\caption{Deadline-bounded and batch scenarios on the real universe, evidence
\texttt{E7}. S4: 25 personalized accounts under a declared 0.5\,s per-account
budget. S5: 200 personalized accounts, queue wall-clock.}
\label{tab:e7batch}
\small
\begin{tabular}{l r r r r r}
\toprule
\hdrrow\hcell{Solver} & \hcell{S4 in-budget} & \hcell{S4 p50 ms} &
\hcell{S5 total s} & \hcell{S5 completed} & \hcell{S5 p99 ms} \\
\midrule
Asymmetry PRISM-CPU & \textbf{25/25} & \textbf{7.4} & \textbf{1.81} & \textbf{200/200} & 42.2 \\
Clarabel & 25/25 & 28.5 & 5.57 & 200/200 & \textbf{34.2} \\
SCS & 24/25 & 36.8 & 9.27 & 175/200 & 128.4 \\
OSQP & 22/25 & 87.2 & 45.34 & 195/200 & 2{,}048.4 \\
Asymmetry PRISM-GPU & 19/25 & 316.5 & 74.69 & \textbf{200/200} & 2{,}849.2 \\
\bottomrule
\end{tabular}
\begin{claimnotes}
Personalized accounts draw randomized holdings drift, return tilts, and
transition penalties around the real data. ``Completed'' counts returns that
pass the external feasibility gates. At this universe size the GPU
configuration's flat call floor exceeds the 0.5\,s budget on 6 of 25
accounts; the stack routes such books to the CPU configuration.
\end{claimnotes}
\end{table}

\begin{figure}[!htb]
\centering
\includegraphics[width=\linewidth]{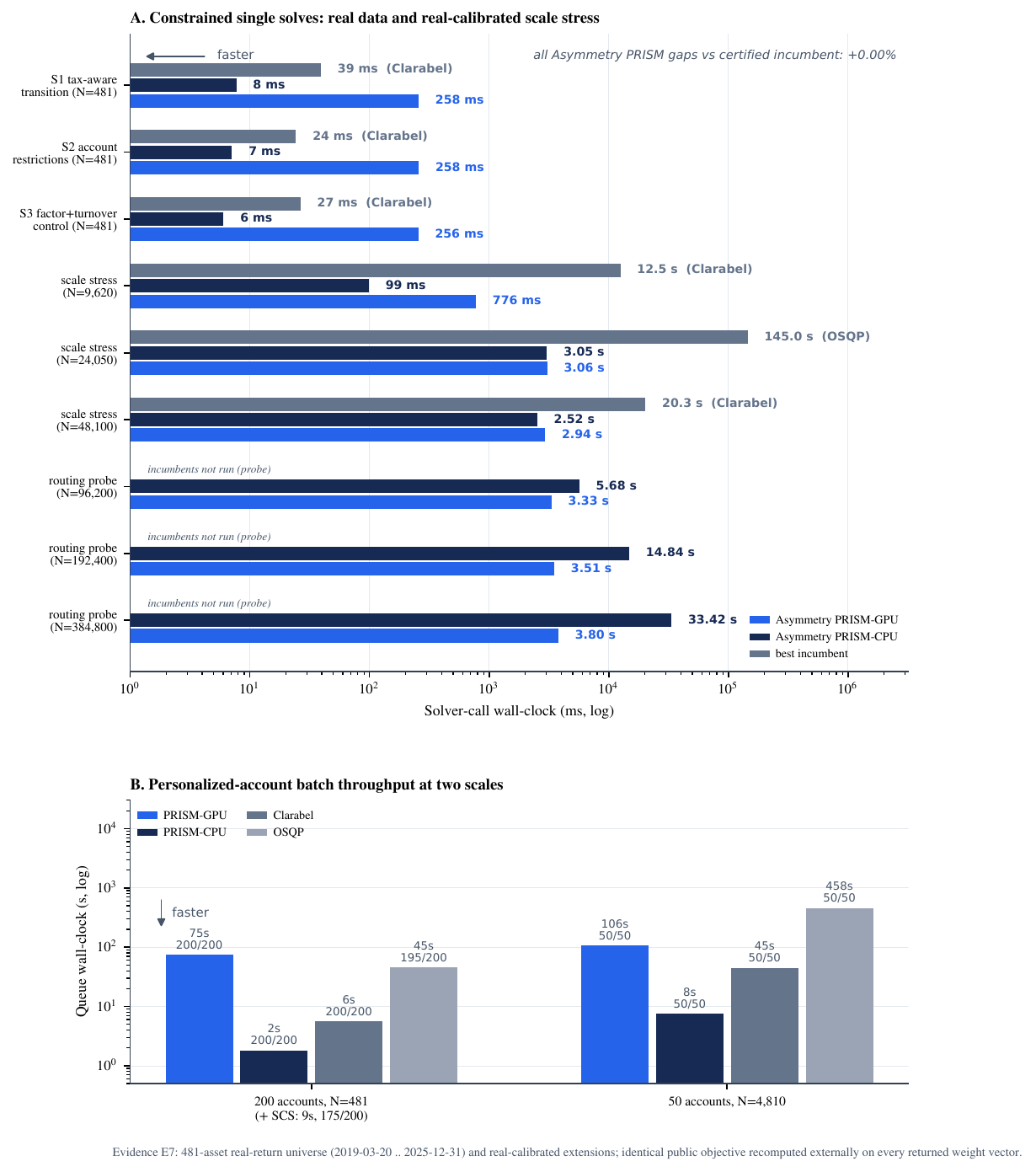}
\caption{Operationally constrained scenarios, evidence \texttt{E7}.
Panel A: solver-call wall-clock on the constrained single solves (real
$N{=}481$ data), the real-calibrated scale-stress rounds with the best
incumbent named per group (Asymmetry PRISM objective gaps versus the certified
incumbent are 0.00\%), and the Asymmetry PRISM-only routing probes, where the GPU
route widens from $1.7\times$ to $8.8\times$ over the CPU route at
identical objectives. Panel B: personalized-account batch throughput at two
scales with gate-passing completion counts.}
\label{fig:sanity}
\end{figure}

The batch result is the operational headline of this section. At
$N{=}481$, Asymmetry PRISM-CPU clears the 200-account book in 1.81\,s with every
account passing the external gates, $3.1\times$ faster than the best
incumbent (Clarabel, 5.57\,s) and $25\times$ faster than OSQP, which leaves
5 accounts unaccepted; SCS leaves 25 of 200 accounts without an accepted
feasible return. A second batch at $N{=}4{,}810$ (50 personalized accounts on
the real-calibrated extension; Figure~\ref{fig:sanity}, Panel~B) repeats the
pattern at scale: Asymmetry PRISM-CPU completes the book in 7.5\,s against 44.7\,s
for Clarabel ($6.0\times$) and 457.6\,s for OSQP ($61\times$), with every
tested path returning 50/50 gate-passing portfolios. In throughput
terms $\Lambda = M/T_{\mathrm{queue}}$, Asymmetry PRISM-CPU sustains
$\Lambda = 200/1.81 = 110.5$ accounts per second at $N{=}481$ and
$\Lambda = 50/7.51 = 6.7$ per second at $N{=}4{,}810$, against $35.9$ and
$1.1$ for the best incumbent. Both Asymmetry PRISM configurations combine full
completion at both scales with certified-equal objectives on the single-solve
scenarios.

\subsection{Scale Stress on a Real-Calibrated Extension}

To test the regime where books are wide as well as deep, the real factor
model is extended to larger cross-sections by resampling factor-loading rows
from the real cross-section and bootstrapping idiosyncratic variances and
return tilts from the real distributions (disclosed as a real-calibrated
extension, not raw constituent data). The same transition objective and lanes
apply.

\begin{table}[!htb]
\centering
\caption{Scale stress on the real-calibrated extension: the S7 study
recorded inside \texttt{E7\_hard\_scenarios\_real\_data}. Solver-call
wall-clock (ms); identical public objective. Where an incumbent ran, Asymmetry PRISM
objectives match the certified incumbent to at least seven decimal places;
the routing-probe rows carry no incumbent comparison.}
\label{tab:e7scale}
\small
\begin{tabular}{r r r r r r r}
\toprule
\hdrrow\hcell{$N$} & \hcell{Asymmetry PRISM-CPU} & \hcell{Asymmetry PRISM-GPU} &
\hcell{Clarabel} & \hcell{OSQP} & \hcell{CPU speedup} & \hcell{GPU speedup} \\
\midrule
1{,}924 & \textbf{58.0} & 1{,}116.0 & 483.7 & 796.7 & 8.3$\times$ & 0.4$\times$ \\
4{,}810 & \textbf{51.6} & 652.0 & 2{,}670.3 & 5{,}747.1 & 51.8$\times$ & 4.1$\times$ \\
9{,}620 & \textbf{98.9} & 776.3 & 12{,}526.5 & 25{,}358.9 & 126.7$\times$ & 16.1$\times$ \\
24{,}050 & \textbf{3{,}050.1} & 3{,}058.7 & 198{,}261.9 & 145{,}046.7 & 47.6$\times$ & 47.4$\times$ \\
48{,}100 & \textbf{2{,}523.5} & 2{,}938.6 & 20{,}269.8 & 489{,}514.2 & 8.0$\times$ & 6.9$\times$ \\
96{,}200\textsuperscript{b} & 5{,}678.0 & \textbf{3{,}329.1} & n.r. & n.r. & n/a & n/a \\
192{,}400\textsuperscript{b} & 14{,}844.4 & \textbf{3{,}506.6} & n.r. & n.r. & n/a & n/a \\
384{,}800\textsuperscript{b} & 33{,}415.6 & \textbf{3{,}795.1} & n.r. & n.r. & n/a & n/a \\
\bottomrule
\end{tabular}
\begin{claimnotes}
Speedups are versus the best completing incumbent at each size (Clarabel at
$N{\leq}9{,}620$ and $N{=}48{,}100$; OSQP at $N{=}24{,}050$). All runs
returned feasible, gate-passing portfolios. Asymmetry PRISM runs at
$N{\geq}24{,}050$ report an iteration-capped status; their externally
recomputed objectives match the certified incumbent to within
$3{\times}10^{-6}$\%, and the two Asymmetry PRISM routes agree to machine precision.
\textsuperscript{b}\,Routing-crossover probes: they locate where the GPU
route overtakes the CPU route and how the margin grows ($1.7\times$,
$4.2\times$, $8.8\times$ at $N{=}96{,}200$, $192{,}400$, $384{,}800$, with
identical objectives at every size); incumbents were not run at these sizes
and no incumbent comparison is claimed.
\end{claimnotes}
\end{table}

Figure~\ref{fig:sanity}, Panel~A places the scale rounds beside the
$N{=}481$ scenarios, and the scaling pattern completes the routing picture of
Section~\ref{sec:solver_results}. Incumbent solve cost is volatile and large
at scale: at $N{=}24{,}050$ the best incumbent needs 145\,s for a single
account, a fifth of the entire 25-minute operating window of
Section~\ref{sec:production}, and at $N{=}48{,}100$ OSQP needs 489.5\,s
while Clarabel recovers to 20.3\,s. Both Asymmetry PRISM routes stay between 2.5\,s
and 3.1\,s across these sizes: a $47.6\times$ advantage at $N{=}24{,}050$ and
$8.0\times$/$6.9\times$ (CPU/GPU) at $N{=}48{,}100$. The GPU route crosses
the best incumbent near $N{\approx}4{,}800$ ($16.1\times$ at $N{=}9{,}620$)
and crosses its own CPU sibling between $N{=}48{,}100$ and $N{=}96{,}200$. From there the
device margin widens with width: $1.7\times$ at $N{=}96{,}200$,
$4.2\times$ at $N{=}192{,}400$, and $8.8\times$ at $N{=}384{,}800$, where
the GPU route returns the identical portfolio in 3.8\,s against 33.4\,s on
CPU. Across a $4\times$ width increase the GPU route moves only from
3.3\,s to 3.8\,s while the CPU route grows roughly linearly, which is the
signature of a configuration whose cost is dominated by parallel arithmetic
rather than problem assembly. Objective agreement holds to at least seven
decimal places at every size, so the scaling advantage carries no quality
concession.

\subsection{Workflow Sanity Checks}

The evidence package also retains walk-forward and Monte Carlo pipeline
checks (\texttt{E3}): repeated portfolio generation over 131 real-data
survivors and 1{,}000 randomized universes, used to verify feasible, stable,
turnover-respecting outputs across rebalances. The recorded pass criteria
are explicit: every rebalance must return weights that satisfy the
Section~\ref{sec:protocol} feasibility gates and the declared turnover cap,
across all 300 walk-forward rebalances and all 1{,}000 Monte Carlo universes;
both checks pass in \texttt{E3}. They are workflow checks under the claim
contract and carry no performance claim.

\section{Evidence Ledger, Artifact Policy, and Claim Notes}
\label{sec:evidence}

This section is the governance layer of the paper. Sections
\ref{sec:solver_results} through \ref{sec:sanity} report what was measured;
this section records where each number lives, what kind of claim it is allowed
to carry, and how an independent reader audits it. The discipline is the same
one applied to the solvers themselves: a claim without a versioned artifact is
treated the way a portfolio without an audit record is treated in
Section~\ref{sec:production}, as operationally unusable.

\subsection{Evidence Artifacts}

\begin{figure}[!htb]
\centering
\includegraphics[width=0.82\linewidth]{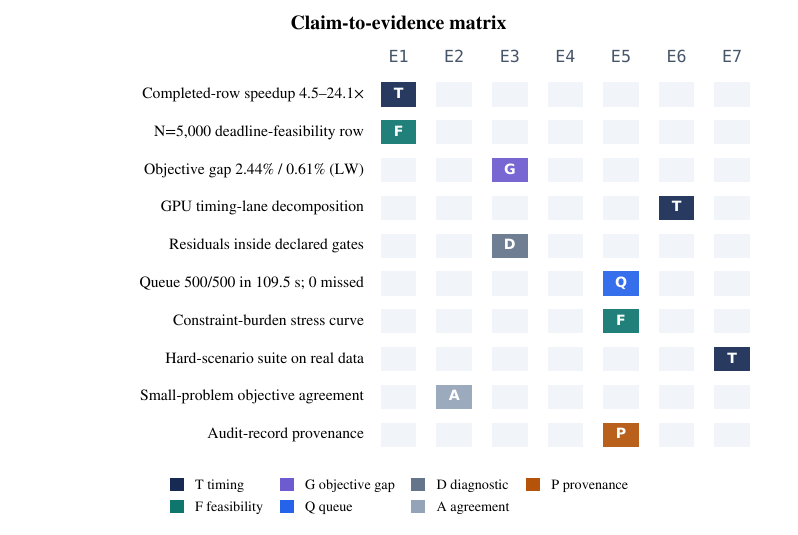}
\caption{Claim-to-evidence matrix. Rows are the central claims of the paper;
columns are the evidence artifacts; cell color encodes the claim type carried
by that artifact under the claim contract of
Table~\ref{tab:claim_contract}.}
\label{fig:claim_matrix}
\end{figure}

Each numerical claim in the paper is tied to one of the evidence artifacts in
Table~\ref{tab:evidence}; Figure~\ref{fig:claim_matrix} maps the central
claims to the artifacts that carry them.

\begin{table}[!htb]
\centering
\caption{Evidence artifacts used in this paper.}
\label{tab:evidence}
\footnotesize
\begin{tabularx}{0.97\linewidth}{l l Y}
\toprule
\hdrrow\hcell{Artifact} & \hcell{Generated} & \hcell{Use in this paper} \\
\midrule
\texttt{E1\_multisolver\_public\_rows} & 2026-05-09 & Multi-solver timings, versions, CPU/GPU timing frontier. \\
\texttt{E2\_small\_problem\_agreement} & 2026-05-09 & Public FF30 small-problem study; comparator objective agreement. \\
\texttt{E3\_external\_diagnostics\_and\_gap\_checks} & 2026-05-09 & KKT-style checks, objective-gap and covariance sensitivity, sanity checks. \\
\texttt{E4\_deadline\_completion\_grid} & 2026-05-09 & Deadline-completion grid, memory-class statement. \\
\texttt{E5\_production\_queue\_500x10000} & 2026-05-09 & Production queue benchmark and constraint-burden curve. \\
\texttt{E6\_gpu\_timing\_lane\_separation} & 2026-06-09 & Supplemental Asymmetry PRISM-GPU timing/status coverage; no objective-gap claims. \\
\texttt{E7\_hard\_scenarios\_real\_data} & 2026-06-11 & Operationally constrained real-data scenario suite and scale stress. \\
\bottomrule
\end{tabularx}
\end{table}

\subsection{Status Vocabulary and Objective Agreement}

\begin{table}[!htb]
\centering
\caption{Status vocabulary used in all result tables.}
\label{tab:statusvocab}
\small
\begin{tabularx}{0.94\linewidth}{l Y}
\toprule
\hdrrow\hcell{Status} & \hcell{Meaning} \\
\midrule
certified & solver reports completion and external checks pass \\
feasible / non-certified (\emph{feas}) & portfolio returned, feasibility checks pass, certification unavailable \\
timeout (\emph{t/o}) & declared wall-clock ceiling reached before return \\
deadline exceeded (\emph{late}) & completed status returned after the operating window \\
no accepted feasible return & no output passed the external gates before the deadline; not a certified infeasibility status \\
allocation / numerical failure & problem object not allocated, or invalid numerical status reported \\
\bottomrule
\end{tabularx}
\end{table}

The small-problem study \texttt{E2\_small\_problem\_agreement} (five seeds,
$N{=}30$) records that all completed comparators agree on the objective value
to approximately seven significant figures, anchoring the objective-agreement
methodology used for gap eligibility. Asymmetry PRISM timing rows are not recorded in
\texttt{E2}; the artifact is comparator-agreement evidence.

\Needspace{10\baselineskip}
\subsection{Artifact Policy}

\begin{reprobox}[title={Artifact note}]
Public claim-bearing artifacts, result tables, environment metadata, and
external validation scripts are archived at
\href{https://github.com/AsymmetryComputing/prism-public-evaluation}{\texttt{github.com/AsymmetryComputing/prism-public-evaluation}}
(release \texttt{v1.0.0}); a DOI-stamped archival snapshot of the same
release accompanies the submission record. External checks on returned weights
are reproducible from the archived data.
\end{reprobox}

\begin{table}[H]
\centering
\caption{Public artifact repository layout.}
\label{tab:repo}
\small
\begin{tabularx}{0.95\linewidth}{l Y}
\toprule
\hdrrow\hcell{Path} & \hcell{Contents} \\
\midrule
\texttt{README.md}, \texttt{MANIFEST.yml}, \texttt{CITATION.cff}, \texttt{LICENSE} & Index, artifact manifest, citation metadata, license. \\
\texttt{environment/} & Hardware/runtime disclosure and package versions. \\
\texttt{data\_sanitized/} & Benchmark input manifest; no client data, no engine state. \\
\texttt{results/} & One sanitized CSV per evidence artifact (\texttt{E1}--\texttt{E7}). \\
\texttt{figures/}, \texttt{tables/} & Published figures (vector) and machine-readable tables. \\
\texttt{validation/} & Public residual and feasibility-gate scripts; table reproduction script. \\
\texttt{evidence\_ledger/} & Claim-to-evidence ledger, artifact manifest, artifact policy. \\
\texttt{paper/} & arXiv version and source. \\
\bottomrule
\end{tabularx}
\end{table}

The reproducibility targets are:

\begin{enumerate}[label=\textbf{R\arabic*.}, leftmargin=2.4em]
  \item every table row has an evidence artifact and generation date;
  \item every speedup compares rows from the same timing lane;
  \item every objective-gap statement identifies a completed reference solver;
  \item every large-scale row without a completed reference avoids certified
        optimality language; and
  \item every production claim includes completion count, deadline status, and
        audit-record coverage.
\end{enumerate}

Two additional non-claim-bearing extensions are recorded in the evidence
ledger: a burden-specific Asymmetry PRISM-GPU rerun of the constraint-burden curve and a
multi-seed rerun of the single-shot \texttt{E1} rows.

\section{Solver Eligibility Ledger}
\label{sec:eligibility}

Table~\ref{tab:versions} records the full eligibility ledger for the evidence
package, including installed packages that did not produce claim-bearing rows.
Entries are eligibility records under the declared protocol.

\begin{table}[!htb]
\centering
\caption{Solver eligibility ledger for the recorded evidence package.}
\label{tab:versions}
\small
\begin{tabularx}{0.97\linewidth}{l l Y}
\toprule
\hdrrow\hcell{Solver stack} & \hcell{Version} & \hcell{Eligibility} \\
\midrule
Asymmetry PRISM-CPU & evidence build \texttt{512758f} & claim-bearing where recorded \\
Asymmetry PRISM-GPU & evidence build \texttt{512758f} & claim-bearing; \texttt{E6} rows are timing/status only \\
OSQP & 1.1.0 & claim-bearing where recorded \\
Clarabel & 0.11.1 & claim-bearing where recorded \\
SCS & via CVXPY 1.8.1 & claim-bearing where recorded \\
MOSEK & 11.1.11 & claim-bearing where recorded \\
CVXPY interface & 1.8.1 & comparator modeling layer; backend named per row \\
Commercial reference (anonymized) & recorded in evidence & objective-gap reference only; timings unpublished under license terms \\
CPLEX & 22.1.2.0 & configured license supported the public-small lane only; not eligible for claim-bearing comparisons \\
HiGHS & 1.13.0 & public-small lane only; not a QP comparator in this package \\
NVIDIA cuOpt & 26.2.0 & no versioned active run recorded; not benchmarked \\
CVXPortfolio & 1.5.1 & own modeling-layer lane; not eligible for same-lane claim-bearing rows \\
\bottomrule
\end{tabularx}
\end{table}

\section{Validity and Threats}
\label{sec:validity}

\textbf{Workload boundary.} Equation~\eqref{eq:public_problem} is a benchmark
interface, not a complete model of every mandate. Discrete lot rules,
market-impact models, client-specific tax instructions, and compliance
overlays require separate protocol rows.

\textbf{Harness asymmetry.} Two disclosed comparator lanes are used: the
\texttt{E1} rows include CVXPY model construction inside the timed call,
while the \texttt{E7} suite pre-builds parametrized CVXPY models and times
only the solver call (the stricter lane for Asymmetry PRISM). Asymmetry PRISM rows are measured
at Asymmetry PRISM's public solver-call boundary in both. The completed-row speedups (up to $24.1\times$) and the queue
outcome (500/500 versus 4/500) far exceed plausible modeling-layer overhead;
millisecond-level differences between harness paths should not be over-read.

\textbf{Objective-gap eligibility.} Gaps are shown only where a reference
solver completed on the same public objective. The gap reference is an
anonymized commercial solver; it contributes completed reference objectives
only. The anonymization is a license constraint and a recognized
reproducibility cost: independent verification of those gaps requires
re-running a licensed commercial solver on the archived instances.

\textbf{Reference incompletion.} A timeout or missing reference is evidence
that the reference row did not complete inside the operating lane under the
recorded protocol, nothing more.

\textbf{Hardware specificity.} Timings are specific to the recorded
workstation and to the disclosed comparator versions; newer point releases of
the comparators may exist, and rows bind to the recorded versions. New
hardware or solver versions require a fresh disclosure table and new evidence
artifacts.

\textbf{Scenario-suite interpretation.} The Section~\ref{sec:sanity} suite
uses real prices and real-return factor models, with account personalization
drawn programmatically around the real data; restriction structures beyond
position caps and L1 transitions require their own protocol rows. The current
S\&P 500 membership implies survivorship in the universe construction.

\textbf{Economic interpretation.} The paper evaluates optimization-system
behavior. Deployments additionally require mandate-specific objectives,
transaction-cost models, tax assumptions, compliance review, and post-trade
monitoring.

\textbf{Timing variation.} Solver timings vary with host load, memory
pressure, driver state, and package versions. The evidence package reports
medians, p99 values, completed counts, and failure rates rather than one-off
best times.

\section{Application Domains and Operating Contexts}
\label{sec:applications}

Institutional rebalancing rarely arrives as a single optimization. It arrives
as a book of accounts, each with its own constraints, that has to be solved
against a fixed cutoff. The benchmark in this paper is built around that shape.
This section connects it to the production settings where the same shape
recurs, and to why those settings have become harder to serve as account
counts grow.

\textbf{Personalized and enhanced direct indexing.}
Direct indexing gives a client direct ownership of a benchmark's constituents
rather than a share in a pooled fund. That ownership is what makes per-client
tax-loss harvesting, wash-sale avoidance, and security-level restrictions
possible, none of which can be expressed at the fund
level~\cite{khang2022,moehle2021taxaware,stein2017}. The value at stake is not
hypothetical: deferring realized gains is a timing option whose worth has been
modeled since the 1980s~\cite{constantinides1983tax}, and a long-run empirical
study puts the tax alpha from systematic loss harvesting on the order of one
percentage point per year before costs~\cite{chaudhuri2020tlh}, which is large
enough that capturing it reliably across an entire book is itself a
reason to reoptimize every account rather than a representative one. The
controls in this benchmark (turnover limits, restriction caps, tax-motivated
transition penalties) are the same controls these platforms apply. What changes at
production scale is the arithmetic: a provider running tens of thousands of
accounts reoptimizes each one on its own cadence, so the account count rather
than the size of the instrument universe sets the compute budget, and the
whole book has to clear before any order goes out. The queue study of
Section~\ref{sec:production}, 500 accounts over a 10,000-instrument universe,
was sized to that regime. As these books grow, the hard part stops being how
well a single account solves and becomes how many accounts close inside the
window.

\textbf{Transition management and model-portfolio distribution.}
A second setting with the same structure is the large portfolio transition:
moving a book between factor allocations, risk budgets, or manager mandates
ahead of a close-related cutoff~\cite{harvey2025rebalancing}. Here the cost of
being slow is not abstract. A portfolio that is half-migrated carries
unintended factor and tracking-error exposure for as long as the transition
runs; this gap between the intended portfolio and the realized one is the
implementation shortfall that has framed execution cost since
Perold~\cite{perold1988shortfall}, and bounding it under a time budget is the
core problem of the optimal-execution literature~\cite{almgren2001execution}.
The practical objective is therefore to compute the full set of target weights
and release the trades inside a single window rather than letting the book
drift across sessions. Transitions also face the same pre-trade risk checks as
routine rebalancing~\cite{sec15c35,esmarts6}, with the added complication that
the starting and target portfolios can differ sharply in exposure, so each
account is optimized against a live target under the clock. The batch and
multi-account scenarios of Section~\ref{sec:sanity} stand in for this: every
account carries the full constraint set (turnover, transition penalty,
restriction cap) and must return a feasible, audited result before the batch
reports completion. The metric that matters is whether the entire batch clears
the window, not whether any one account solves quickly, which is why the
results in this paper report completed-account counts alongside per-solve
times.

\textbf{Regulatory and audit obligations.}
Pre-trade controls under SEC Rule 15c3-5, MiFID II RTS\,6, and the IOSCO
principles for direct electronic access require a completed, bounded risk check
before an order reaches the market~\cite{sec15c35,esmarts6,iosco2010dea}. When
the order originates from an optimizer, the solve sits inside that gate:
nothing trades until a feasible weight vector with a checkable quality bound
exists, so a solver that stalls or returns an unverifiable result does not just
slow the pipeline, it blocks order flow entirely. This is the reason the
benchmark treats feasibility status and an external quality bound as
first-class outputs rather than diagnostics. The per-solve audit record of
Section~\ref{sec:production}, recording input identity, output identity,
feasibility status, and elapsed time, is the computational side of the same
obligation: it is what lets a supervisor reconstruct, after the fact, exactly
which inputs produced which trade and whether the deadline was met. The number
of such records grows one-for-one with the account count, which is a further
reason to close accounts in parallel inside the window rather than one after
another against it.

\textbf{Outlook.}
Three trends are making this workload both more common and more demanding at
once. Account counts are rising as direct indexing moves down-market from
ultra-high-net-worth mandates toward retail platforms; personalization is
deepening, with per-client factor models and restriction sets replacing a
single shared model~\cite{fama1993common,ledoitwolf2004}; and the numerical
stack is shifting onto GPUs, where large quadratic programs admit data-parallel
solution~\cite{schubiger2020gpu} and throughput is gained across many instances
at once rather than by shortening any single solve. Together these move the
binding cost from how well one instance solves to how a whole fleet is
scheduled before a deadline, a regime in which classical per-instance solver
benchmarks no longer predict production behavior. Recent work on cost-aware
optimization over large universes~\cite{fan2025costaware} and on
GPU-accelerated portfolio solvers~\cite{flashfolio2026,niu2026scalable} is
moving in the same direction, and the evidence here adds one measured point to
that line: under the recorded protocol the engine cleared a 500-account
production queue in 109.5\,s, against 4 of 500 for the recorded baseline in the
same window. The reader's takeaway is the framing as much as the number. For
this class of workload the deadline is the binding constraint rather than a
side condition, throughput is measured per book and not per solve, and a
benchmark that does not report completed-account counts under a declared window
is not measuring the quantity the operator actually cares about.

\section{Conclusion}
\label{sec:conclusion}

Asymmetry PRISM is evaluated as a CPU/GPU portfolio optimization engine for
deadline-bounded institutional rebalancing, through a public evaluation
boundary: timings, feasibility checks, eligible objective gaps, KKT-style
diagnostics, memory class, failure rates, queue completion, and auditability.
On the recorded evidence, Asymmetry PRISM-CPU is $4.5\times$ to $24.1\times$ faster than
the fastest completed reference rows at $N{=}100$ to $N{=}2{,}000$ and returns
externally feasible portfolios with bounded reference gaps where eligibility
holds. On the real-data scenario suite, Asymmetry PRISM clears transition-penalized,
restriction-capped, turnover-controlled, and scale-stressed solves
$3.4\times$ to $126.7\times$ faster than the best completing incumbent at
certified-equal objectives, completes the 200-account batch in 1.81\,s with
every account passing the external gates, and the GPU route widens to
$8.8\times$ over the CPU route at $N{=}384{,}800$. The headline result is systems-level:
under the recorded protocol, Asymmetry PRISM-GPU completed the full 500-account,
10,000-instrument production queue in 109.5\,s within the declared 25-minute
window, with zero missed deadlines and an audit record for every solve, while
the recorded OSQP queue baseline completed 4 of 500 accounts. Every number in
those statements traces to the evidence artifacts of
Section~\ref{sec:evidence}.

\bibliographystyle{plainnat}
\bibliography{mybib}

\end{document}